\documentclass[11pt]{article}
\usepackage{times,graphicx,theorem,epsfig,lscape,dcolumn,multirow,epic,float}
\usepackage{amsmath,amssymb}
\usepackage[normalem]{ulem}
\usepackage{subfigure}
\usepackage{adjustbox}
\usepackage{natbib}
\usepackage[figuresright]{rotating}
\usepackage[english]{babel}
\usepackage{booktabs}
\usepackage{hhline}
\usepackage{bm}
\usepackage{enumerate}
\usepackage{verbatim}
\usepackage{dsfont}
\usepackage[colorlinks=true,linkcolor=black,citecolor=black]{hyperref}
\usepackage{threeparttable}
\usepackage{array}
\usepackage{comment}
\usepackage{pdflscape}
\usepackage{afterpage}
\usepackage{booktabs}
\usepackage{moreverb}
\usepackage{graphicx}
\usepackage{tikz}
\usepackage[labelsep=quad,skip=-2pt]{caption}

\numberwithin{equation}{section}
\theoremstyle{plain} 
\theoremstyle{plain} 
\theoremstyle{plain} 
\theoremstyle{plain} 
\theoremstyle{plain} 

\theoremstyle{plain}

\theoremstyle{plain} 

\theoremstyle{plain}

\newcommand{\blinding}[2]{#1}   

\newcommand{\bX}{\mbox{\bf X}}
\newcommand{\bx}{\mbox{\bf x}}

\newcommand{\bxi}{\boldsymbol{\xi}}
\newcommand{\bgamma}{\boldsymbol{\gamma}}
\newcommand{\bbeta}{\boldsymbol{\beta}}

\newcommand{\be}{\begin{eqnarray}}
	\newcommand{\ee}{\end{eqnarray}}
\newcommand{\bee}{\begin{eqnarray*}}
	\newcommand{\eee}{\end{eqnarray*}}
\newcommand{\bi}{\begin{enumerate}[(i)]}
	\newcommand{\ei}{\end{enumerate}}

\def\checkmark{\tikz\fill[scale=0.4](0,.35) -- (.25,0) -- (1,.7) -- (.25,.15) -- cycle;} 

\usepackage{tikz-cd}
\tikzset{smalltext/.style={"\textup{\small #1}" description}}

\begin{document}

\begin{center}
\vspace*{-2.5cm}
{\Large Assessing mediation in cross-sectional stepped wedge cluster randomized trials}

\medskip
\blinding{
Zhiqiang Cao$^{1}$ and Fan Li$^{2,3,4,*}$
}{}

$^{1}$Department of Mathematics, College of Big Data and Internet, Shenzhen Technology University, Guangdong, China. \\
$^{2}$Department of Biostatistics, Yale School of Public Health, New Haven, Connecticut, USA. \\
$^{3}$Center for Methods in Implementation and Prevention Science, Yale School of Public Health, New Haven, Connecticut, USA.\\
$^{4}$Clinical and Translational Research Accelerator, Department of Medicine, Yale School of Medicine, New Haven, Connecticut, USA.\\
$*$fan.f.li@yale.edu\\

\end{center}

\date{}

{\centerline{Abstract}
\noindent Mediation analysis has been comprehensively studied for independent data but relatively little work has been done for correlated data, especially for the increasingly adopted stepped wedge cluster randomized trials (SW-CRTs). Motivated by challenges in understanding the effect mechanisms in pragmatic and implementation science clinical trials, we develop new methods for mediation analysis in SW-CRTs. Specifically, based on a linear and generalized linear mixed models, we demonstrate how to estimate the natural indirect effect and mediation proportion in typical SW-CRTs with four data types, including both continuous and binary mediators and outcomes. Furthermore, to address the emerging challenges in exposure-time treatment effect heterogeneity, we derive the mediation expressions in SW-CRTs when the total effect varies as a function of the exposure time. The cluster jackknife approach is considered for inference across all data types and treatment effect structures. We conduct extensive simulations to evaluate the finite-sample performance of the proposed mediation estimators and demonstrate the proposed approach in a real data example. An R package \texttt{mediateSWCRT} has been developed to facilitate the practical implementation of the estimators. 
\vspace*{0.3cm}

\noindent {\sc Key words}: Mediation analysis, Stepped wedge cluster randomized trials, Time-dependent treatment effect, Natural
indirect effect; Mediation proportion; Jackknife variance
}

\clearpage

\section{Introduction}
\label{sec1}

A cluster randomized trial (CRT) is a type of randomized controlled experiment in which pre-existing groups, called clusters, of individuals are randomly allocated to treatment arms \citep{Murray1998,Donner2000}.  
CRTs can be used when individual randomization to treatment arms is not feasible due to logistical or administrative considerations, when the intervention is naturally administered on a community-wide scale \citep{Gail1996}, or when there is a desire to reduce the chance of treatment contamination \citep{Torgerson2001}. From a statistical standpoint, a key characteristic of CRTs is that the individual outcomes within a cluster are correlated and this intracluster correlation coefficient (ICC) must be incorporated into power calculation and the trial analysis; see, for example, a pair of methodological review of methods for CRTs in \citet{Turner2017a,Turner2017b}. CRTs often employ a parallel-arm design, that is, for a two-arm study with $2I$ independent clusters, $I$ clusters are randomly
assigned to each intervention at a single time point. 
An increasingly popular CRT design is the stepped wedge cluster randomized trial (SW-CRT), which randomly allocates clusters into uni-directional and distinct treatment sequences that are marked by the treatment adoption time. 
That is, in a SW-CRT, all clusters start from the control condition and 
a subset of clusters are randomly selected to cross-over to the intervention condition at pre-determined time points in a staggered fashion, until all clusters receive the intervention \citep{Hussey2007,Hemming2015b}.
\citet{Hemming2020} provided a comprehensive discussion on when stepped wedge design is a good study design choice for 
evaluating the population-level impact of an intervention. Methodological review of available statistical methods for designing and analyzing SW-CRTs can be found in \citet{Li2021} and \citet{li2022stepped}.

While the total treatment effect has historically been the primary focus in analyzing CRTs, there is emerging interest, for example, under the hybrid type I-III implementation study framework \citep{curran2012effectiveness}, in addressing the mechanisms by which a cluster-level intervention improves individual outcomes, and in elucidating the key pathways for optimizing treatments leading to the better health outcomes \citep{williams2016multilevel,lewis2020systematic}. Mediation analysis methods are quantitative statistical methods designed for this objective in general, and findings from mediation analyses hold the promise to inform healthcare policy, optimize system-level interventions, and facilitate the uptake of policies into clinical practice. Methods for mediation analysis have been extensively studied for independent data \citep{VanderWeele2009,VanderWeele2015,vanderweele2016mediation}, accommodating different types of outcomes and mediators \citep{Nevo2017,Cheng2021,cheng2023product}, and there are also efforts that expand the mediation methods to applications in CRTs. For example, several studies \citep{krull1999multilevel,bauer2006conceptualizing,zhang2009testing} have considered linear mixed models to appropriately represent the correlated mediators and outcomes within each cluster and developed estimators and tests for the natural indirect effect in CRTs. \citet{park2015bayesian} studied identification conditions for the natural indirect effect in CRTs, and developed a Bayesian multilevel modeling approach for assessing mediation. \citet{vanderweele2013mediation} considered the possibility of within-cluster interference and studied nonparametric identification for both the natural indirect effect and the spillover mediation effect in CRTs. In a similar setting, \citet{cheng2024semiparametric} developed the semiparametric efficiency theory and formalized semiparametric efficient estimators for natural indirect effect and the spillover mediation effect, allowing for both parametric and nonparametric working models. These prior efforts, however, have solely focused on the parallel-arm designs and have not addressed the unique complexities in stepped wedge designs due to staggered treatment allocation. While we are unaware of a previous study that investigates mediation methods in SW-CRTs, there have already been several published mediation analyses of completed SW-CRTs \citep{stevens2019mechanisms,gosselin2023immigrants} and protocols that plan for mediation analyses in ongoing SW-CRTs \citep{suresh2022pathweigh,shelley2015testing}. Thus, it remains of substantial interest to formally investigate extensions of mediation methods to SW-CRTs with different data types for guiding practice.

In this article, we provide a systematic investigation of regression-based mediation analysis of cross-sectional SW-CRTs (the most common type of stepped wedge design according to an empirical review by \citet{nevins2024adherence}), where different individuals are included in different study periods. We consider four common data types, including a continuous outcome with a continuous mediator ($Y_cM_c$), a continuous outcome with a binary mediator ($Y_cM_b$), a binary outcome and a continuous mediator ($Y_bM_c$), and finally, a binary outcome with a binary mediator ($Y_bM_b$). For each data type, we describe the linear mixed model or generalized linear mixed model for the outcome and mediator, and develop the corresponding expressions of three mediation effect measures, that is, the {\it natural indirect effect} (NIE) describing the effect through the mediator, the {\it natural direct effect} (NDE) describing the effect around the mediator and solely due to the treatment, as well as the {\it mediation proportion} (MP) defined as the ratio of NIE and total effect (TE, or the sum of NIE and NDE). We start with an immediate and sustained treatment effect model as in \citet{Hussey2007}, and then further generalize our derivations to accommodate the exposure-time treatment effect model that allows the treatment effect to change by the duration of treatment. Several recent studies \citep{Kenny2022,Maleyeff2023,wang2024achieve} have highlighted the importance of addressing the exposure-time treatment effect heterogeneity when it exists to avoid potentially substantial bias in estimating the total treatment effect in SW-CRTs. As the total treatment effect consists of the basis for decomposition into NIE and NDE, the importance of addressing exposure-time treatment effect heterogeneity carries over to mediation analysis of SW-CRTs. For each data type and treatment effect structure, we articulate expressions for point estimation and pursue inference via a cluster-jackknife approach. Finally, we provide a new R package, \texttt{mediateSWCRT} to implement our proposed methods and facilitate mediation analysis in SW-CRTs. 

The remainder of this article is organized as follows. In Section \ref{sec:notation}, we introduce notation and assumptions for mediation analysis of SW-CRTs, and define the mediation effect measures of interest in this work. We develop estimators for mediation effect measures in SW-CRTs under an instantaneous and sustained treatment effect assumption with each data type in Section \ref{sec:const_te}. In Section \ref{sec:td_te}, we move onto extending the mediation effect measures to SW-CRTs in the presence of an exposure-time dependent treatment effect structure. We then describe a simulation study to evaluate the finite-sample performance of the proposed mediation methods under an instantaneous and sustained treatment effect assumption in Section \ref{sec:simu_constant}. To further illustrate our methods, we re-analyze a pragmatic, cross-sectional SW-CRT in Uganda via mediation analysis by assuming an instantaneous and constant treatment effect structure and a time-dependent treatment effect structure in Section \ref{sec:application}. Concluding remarks and discussions for future research are provided in Section \ref{sec:discussion}.

\section{Notation and Models for Mediation Analysis in Stepped Wedge Designs}\label{sec:notation}
\begin{figure}
\centering
\subfigure[A schematic of a stepped wedge design with $I=8$ clusters and $J=5$ periods.]{\includegraphics[width=5in]{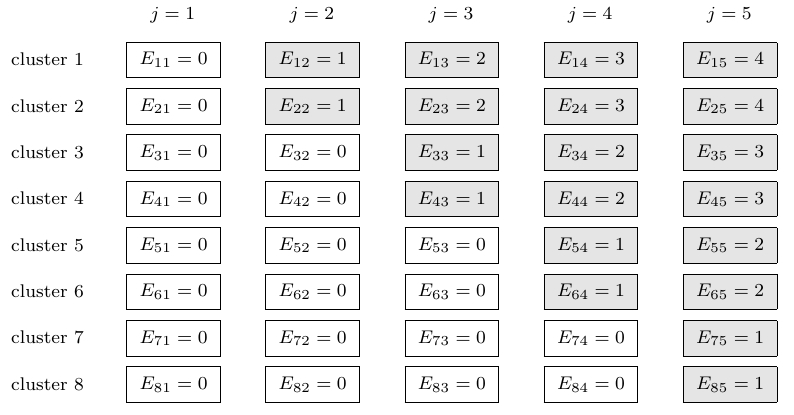}} 
\\
\centering
\subfigure[Illustrative mediation directed acyclic graph in cluster $1$ during periods $j=1$ and $j=2$.]{\includegraphics[width=5in]{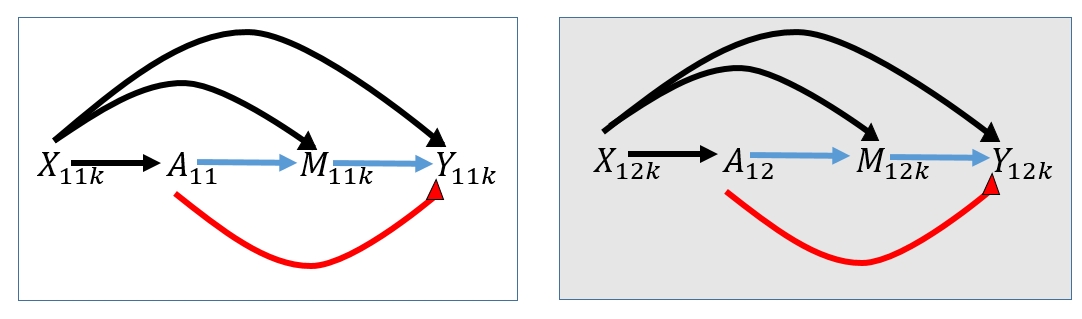}}
\caption{Panel (a): A schematic illustration of a stepped wedge design with $I=8$ clusters and $J=5$ periods, with two clusters randomized to each distinct treatment sequence. Each white cell indicates a control ($A_{ij}=0$) cluster-period and each grey cell indicates a treatment ($A_{ij}=1$) cluster-period. To explain the concept of exposure time, the value of $E_{ij}$---the duration of treatment---is included in each cluster-period.  Panel (b): An example mediation directed acyclic graph in cluster $1$ during periods $j=1$ and $j=2$, where $Y_{11k}$ (and $Y_{12k}$), $A_{11}$ (and $A_{12}$), $M_{11k}$ (and $M_{12k}$) and $X_{11k}$ (and $X_{12k}$) denote individual $k$'s outcome, exposure, mediator, and baseline confounders of the exposure-outcome
and exposure-mediator relationships, with $k \in \{1,\dots,N_{11}\}$ (or $k \in \{1,\dots,N_{12}\}$). }\label{fig:scheme}
\end{figure}

We consider SW-CRTs with $I$ clusters followed over $J$  ($J \ge 3$) time periods. In this design, we assume $E$ is the maximum exposure time periods, or equivalently, the maximum number of periods a cluster can be exposed under the intervention condition during the trial. At the cluster-period level, we let $E_{ij}\in\{0,1,\ldots,E\}$ be the exposure time for cluster $i$ ($i\in\{1,\ldots,I\}$) during period $j$ ($j=1,\ldots,J$), and therefore $A_{ij}=\mathbb{I}(E_{ij}>0)$ denotes the binary treatment status of cluster $i$ during period $j$. Typically, the first period is a baseline period such that no clusters receive treatment with $A_{i1}=0$ for all $i$. In each post-baseline period, a subset of clusters may be randomly selected to transition from control to treatment, until all clusters are exposed under treatment such that $A_{iJ}=1$ for each $i$. We focus on a cross-sectional design where distinct individuals are enrolled in different time periods, and use $N_{ij}$ to denote the cluster-period size; that is, the number of individuals accessed during period $j$ in cluster $i$. For simplicity, we assume that $N_{ij}$ is only randomly varying, and therefore do not further address informative cluster-period sizes \citep{kahan2023estimands}; discussions of estimands under informative cluster-period sizes can be found in \citet{chen2023model}.
At the individual level, we denote $Y_{ijk}$ to be the observed final outcome for individual $k$ in cluster $i$ during period $j$, which can be either continuous or binary. Prior to the observation of the final outcome but post treatment assignment, we consider an intermediate outcome $M_{ijk}$ (which can be either continuous or binary) as a potential mediator that lies in the causal pathway between $A_{ij}$ and $Y_{ijk}$. 
In addition, we assume the access to a vector of baseline covariates $\bX_{ijk}$. Here we consider $\bX_{ijk}$ to be measured just prior to the enrollment of individual $k$ in cluster $i$ during period $j$ and thus not affected by treatment assignment $A_{ij}$ nor any prior treatment history at the cluster-level; in other words, $\bX_{ijk}$ is exogenous. For notation brevity, we also consider $\bX_{ijk}$ to include all time-invariant cluster-level covariates such as geographical location that may be potential confounders. 
A schematic illustration of a standard stepped wedge design with $I=8$ clusters and $J=5$ periods is shown in panel (a) of Figure \ref{fig:scheme}. Zooming into the first two periods in cluster 1, panel (b) of Figure \ref{fig:scheme} presents a directed acyclic graph illustrating the causal relationship between the observed variables for two randomly selected individuals. 

\begin{center}
[Figure \ref{fig:scheme} about here.]
\end{center}

To assess mediation in a SW-CRT, we \textcolor{black}{primarily} consider the following paired generalized linear mixed models for the outcome $Y_{ijk}$ and the mediator $M_{ijk}$ respectively, that is
\begin{eqnarray}\label{outcome_m}
	g\left(E(Y_{ijk}|A_{ij},M_{ijk},\bX_{ijk},\alpha_i)\right)=\beta_{0j}+\theta(E_{ij})A_{ij}+\beta_{M}M_{ijk}+\bbeta_X^T\bX_{ijk}+\alpha_i,
\end{eqnarray}
and
\begin{eqnarray}\label{mediator_m}
	h\left(E(M_{ijk}|A_{ij},\bX_{ijk},\tau_i)\right)=\gamma_{0j}+\eta(E_{ij})A_{ij}+\bgamma_X^T\bX_{ijk}+\tau_i,
\end{eqnarray}
where both $g(\cdot)$ and $h(\cdot)$ are link functions, and to reflect the conventional practice in SW-CRTs, we consider the identity link function for continuous variables and logistic link function for binary variables. 
In the above models, $\beta_{0j}$ and $\gamma_{0j}$ represent the underling secular trend or period effect; $\theta(E_{ij})$ and $\eta(E_{ij})$ represent the treatment effect parameters that may potentially vary as a function of the exposure time (duration of treatment) for the outcome and mediator. 
Under the assumption of an instantaneous and constant treatment effect on both $Y_{ijk}$ and $M_{ijk}$, we set $\theta(E_{ij})=\theta$ and $\eta(E_{ij})=\eta$, and models \eqref{outcome_m}-\eqref{mediator_m} resemble the conventional model considered in \citet{Hussey2007}. Relaxing this assumption, one can model the treatment effects as functions of the exposure time $E_{ij}$ such that $\theta(E_{ij})=\theta_{E_{ij}}$ and $\eta(E_{ij})=\eta_{E_{ij}}$\citep{Hughes2015}. This approach addresses exposure-time treatment effect heterogeneity, which can occur, for example, when there is a learning effect or weakening effect of treatment over time. Several recent studies have emphasized the importance of modeling exposure-time treatment effect heterogeneity when it exists to avoid substantial bias in inference \citep{Kenny2022,Maleyeff2023,wang2024achieve}, and this implication carries to mediation analysis. 
In model \eqref{outcome_m}, $\beta_M$ is the coefficient of the mediator and $\bbeta_X$ represents the confounder effect. Similarly, $\bgamma_X$ is the confounder effect in the mediator model. 
In both models, we consider a cluster-level random intercept to account for correlated outcomes within clusters, given by $\alpha_i \sim N(0,\sigma^2_{\alpha})$ and $\tau_i \sim N(0,\sigma^2_{\tau})$; these two random effects are also assumed mutually independent such that there is no unmeasured cluster-level confounding. Without loss of generality, we also assume both the outcome model and mediator model share the same set of covariates $\bX_{ijk}$. If certain covariates in the outcome and mediator models are considered irrelevant based on content knowledge, we can simply set the corresponding elements in $\bbeta_X$ and $\bgamma_X$ to zero and our results can still apply. \textcolor{black}{Finally, although there have been developments on more complex random-effects assumptions for SW-CRTs (see \citet{Li2021} for a review of model variants), we start with the random-intercept models because (1) they are easy to implement in standard software and less likely to encounter non-convergence issues, and (2) the simplicity of the random-effects structure simplifies the presentation and is helpful to deliver the core idea of the methodology. It is possible to extend our methods to allow for more complex random-effects structures. As a concrete example, we provide an extension of our methods under the nested exchangeable random-effects structure in Web Appendix C, and implement this extension in our R package. A further discussion can be found in Section \ref{sec:discussion}.}

Because the notation for mediation analysis under different treatment effect structures may be different, we first focus on introducing the mediation effect measures, assumptions, and methods for estimation in Section \ref{sec:const_te} before transitioning to the exposure-time treatment effect models in Section \ref{sec:td_te}.

\section{Mediation analysis under an instantaneous and constant treatment effect structure}\label{sec:const_te}
\subsection{Mediation effect measures and assumptions}\label{sec:assumption1}
We start with the instantaneous and constant treatment effect structure, in which case the paired generalized linear mixed models for the outcome and mediator become
\begin{eqnarray}\label{const_models}
	g\left(E(Y_{ijk}|A_{ij},M_{ijk},\bX_{ijk},\alpha_i)\right)&=&\beta_{0j}+\theta A_{ij}+\beta_{M}M_{ijk}+\bbeta_X^T\bX_{ijk}+\alpha_i,\nonumber\\
	h\left(E(M_{ijk}|A_{ij},\bX_{ijk},\tau_i)\right)&=&\gamma_{0j}+\eta A_{ij}+\bgamma_X^T\bX_{ijk}+\tau_i.
\end{eqnarray}
Under these two regression models, we are interested in three mediation effect measures, referred to as the {\it natural indirect effect} (NIE), the {\it natural direct effect} (NDE) as well as the {\it mediation proportion} (MP) defined as the ratio of NIE and total effect (TE, defined as $\text{NIE}+\text{NDE}$). To properly define the mediation effect measures, we consider the potential outcomes notation. Because the treatment effect structure is not a function of the exposure time, we can define $Y_{ijk}(a)$ and $M_{ijk}(a)$ as the potential outcome and mediator for each individual when setting $A_{ij}=a\in\{0,1\}$ at the cluster-period level. Similarly, we denote $Y_{ijk}(a,m)$ as the potential outcome $Y_{ijk}$ when setting $A_{ij}=a$ and $M_{ijk}=m$. 
Then, conditional on the baseline covariate values, $\bX_{ijk}=\bx$, when the treatment level is changed from control ($a=0$) to intervention ($a=1$), the NIE and NDE at calendar time period $j$ can be defined on the $g$-function scale \citep{Nevo2017,cheng2023product}:
\begin{eqnarray}\label{ave_def}
	\text{NIE}(j|\bx)&=&g\left(E[Y_{ijk}(1,M_{ijk}(1))|\bX_{ijk}=\bx]\right)-g\left(E[Y_{ijk}(1,M_{ijk}(0))|\bX_{ijk}=\bx]\right),\nonumber \\
	\text{NDE}(j|\bx)&=&g\left(E[Y_{ijk}(1,M_{ijk}(0))|\bX_{ijk}=\bx]\right)-g\left(E[Y_{ijk}(0,M_{ijk}(0))|\bX_{ijk}=\bx]\right).
\end{eqnarray}
Therefore, the MP during period $j$ is given by the ratio, $\text{MP}(j|\bx)=\text{NIE}(j|\bx)/\text{TE}(j|\bx)=\text{NIE}(j|\bx)/\{\text{NIE}(j|\bx)+\text{NDE}(j|\bx)\}$, with its plausible range as the unit interval $[0,1]$. 
Although it is helpful to investigate the mediation effect measures during each calendar period separately, investigators are frequently interested in a single summary of mediation effect and hence we further define an overall summary of the NIE and NDE based on a calendar period average, given by 
\begin{eqnarray}\label{ave_def_total}
	\text{NIE}(\bx)&=&\sum_{j=1}^J \omega(j)\text{NIE}(j|\bx),~~~
	\text{NDE}(\bx)=\sum_{j=1}^J\omega(j)\text{NDE}(j|\bx),\nonumber \\
	\text{MP}(\bx)&=&\sum_{j=1}^J \left\{\frac{\text{NIE}(j|\bx)+\text{NDE}(j|\bx)}{\sum_{l=1}^J\text{NIE}(l|\bx)+\text{NDE}(l|\bx)}\right\}\text{MP}(j|\bx).
\end{eqnarray}
In what follows, we primarily focus on the uniform weight $\omega(j)=1/J$ for simplicity and interpretability. We note, however, there can be different possibilities in specifying the weights $\omega(j)$ to flexibly determine the relative contribution from each period to the overall summary (for example, a precision weight based on inverse variance may be possible but is arguably less interpretable), and this is a subject for future research. In addition, the mediation proportion $\text{MP}(\bx)$ is invariant to the choice of weight. Of note, the mediation effect measures we pursue in this article are marginal with respect to the clusters (therefore not conditional on any random effects) but are fixing the confounders at the level of $\bX=\bx$. This is standard in regression-based mediation analysis with independent data; see, for example, \citet{vanderweele2016mediation}, \citet{Nevo2017}, \citet{cheng2023product} and we adapt these definitions to the context of stepped wedge designs. 
To apply these definitions, a common practice is to set $\bx$ to be the median levels of $\bX_{ijk}$ \citep{cheng2023product}, as the scale and location of the confounder variables generally will not affect the estimation of the mediation effect measures \citep{Li2007}.

Both the NIE and NDE parameters in (\ref{ave_def}) are not identifiable without additional structural assumptions that can be explicitly stated with the potential outcomes notation. We give these assumptions below under the instantaneous and sustained treatment effect structure. They include (A.1) when we set $A_{ij}=a\in\{0,1\}$, the observed outcome and mediator become $Y_{ijk}=Y_{ijk}(a)$ and $M_{ijk}=M_{ijk}(a)$; furthermore, the observed outcome $Y_{ijk}=Y_{ijk}(a,m)$ when we set $A_{ij}=a$ and $M_{ijk}=m$; (A.2) the potential outcome under $A_{ij}=a$ is equal to the potential outcome when $A_{ij}=a$ and $M_{ijk}$ is set to its natural value under $A_{ij}=a$, that is, $Y_{ijk}(a)=Y_{ijk}(a,M_{ijk}(a))$; (A.3) (i) $Y_{ijk}(a,m) \perp A_{ij}|\bX_{ijk}$; (ii) $Y_{ijk}(a,m) \perp M_{ijk}|(\bX_{ijk},A_{ij})$; (iii) $M_{ijk}(a) \perp A_{ij}|\bX_{ijk}$, where $\perp$ refers to independence; (A.4) For all $a$, $a^*$ and $m$, $Y_{ijk}(a,m)\perp M_{ijk}(a^*)|\bX_{ijk}$. Among them, Assumption (A.1) is the causal consistency assumption, which requires that the treatment and mediator are defined and measured unambiguously, and rules out individual-level interference. It also assumes away any potential dependence of the outcome and mediator on the prior treatment history of a cluster, given the current treatment status $A_{ij}=a$. Assumption (A.2) is the composition assumption \citep{VanderWeele2009}. Assumption (A.3) is the conditional independence assumption \citep{VanderWeele2015}, and parts (i),(ii) and (iii) assume that the treatment-outcome, mediator-outcome and treatment-mediator relationships are unconfounded given the measured covariates $\bX_{ijk}$. Importantly, part (i) and part (iii) under (A.3) hold automatically in SW-CRTs because the clusters are randomized to distinct treatment sequences. Therefore, within each period $j$, $A_{ij}$ is marginalized randomized and is independent of any post-randomization variables. In other words, we only require the collection of measured covariates to deconfound the mediator-outcome relationship. Finally, (A.4) is sometimes termed cross-world independence assumption, which assumes that none of the confounders in the mediator-outcome relationship can be affected by treatment $A_{ij}$, or there is no treatment-induced confounding \citep{Doretti2022}.

Under these structural assumptions and the paired generalized linear mixed models \eqref{const_models}, we next derive explicit expressions of the mediation effect measures, by considering four different data types, including a continuous outcome with a continuous mediator ($Y_cM_c$), a continuous outcome with a binary mediator ($Y_cM_b$), a binary outcome and a continuous mediator ($Y_bM_c$), and finally, a binary outcome with a binary mediator ($Y_bM_b$).

\subsection{Data type 1---A continuous outcome and a continuous mediator ($Y_cM_c$)}\label{sec:cc_hhm}
When both the outcome and mediator are continuous variables, we set $g$ and $h$ to be the identity functions, and specifically write the outcome model (\ref{outcome_m}) and the mediator model (\ref{mediator_m}) as a pair of linear mixed models
\begin{eqnarray}
	Y_{ijk}&=&\beta_{0j}+\theta A_{ij}+\beta_{M}M_{ijk}+\bbeta_X^T\bX_{ijk}+\alpha_i+\epsilon_{ijk}, \label{cont_Y}\\
	M_{ijk}&=&\gamma_{0j}+\eta A_{ij}+\bgamma_{X}^T\bX_{ijk}+\tau_i+e_{ijk},\label{cont_M}
\end{eqnarray}
where, in addition to the random intercepts, $\epsilon_{ijk}\sim N(0,\sigma^2_{\epsilon})$ and $e_{ijk} \sim N(0,\sigma^2_{e})$ are independent residual error terms in the outcome model and mediator model, respectively. 
By the definitions of mediation effect measures in (\ref{ave_def}), we obtain from models \eqref{cont_Y} and \eqref{cont_M} that
\begin{eqnarray*}
	\text{NIE}(j|\bx)&=&\beta_{M}\eta ,\\
	\text{NDE}(j|\bx)&=&\theta,
\end{eqnarray*}
where the detailed derivation is given in Web Appendix A1. Given that both $\text{NIE}(j|\bx)$ and $\text{NDE}(j|\bx)$ do not involve $j$ (owing to the linear mixed model structure), the expressions for NIE and NDE based on summary \eqref{ave_def_total} are equal to $\beta_{M}\eta$ and $\theta$, respectively. These mediation effect measures are also free of covariates, resembling the classical Baron and Kenny result for independent data \citep{baron1986moderator}. Furthermore, the MP is given by $ \frac{\beta_{M}\eta}{\beta_{M}\eta+\theta}$. These results are also consistent with those under the case 1 in \citet{Cheng2021} derived based on linear models and independent data, and present a typical feature of the product of coefficients (i.e., the NIE expression). To obtain NIE, NDE and MP, one can fit models \eqref{cont_Y} and \eqref{cont_M} using maximum likelihood or restricted maximum likelihood approaches (e.g. with the \texttt{nlme} or \texttt{lme4} R package), and plug in the coefficient estimators to the derived NIE and NDE expressions above. 

\subsection{Data type 2---A continuous outcome and a binary mediator ($Y_cM_b$)}\label{sec:cb_hhm}
In this scenario, the outcome model of $Y_{ijk}$ remains the same as model (\ref{cont_Y}), but we model the binary mediator using the following logistic generalized linear mixed model, that is,
\begin{eqnarray}\label{bina_M}
	{\rm logit}\left(P(M_{ijk}=1|A_{ij},\bX_{ijk},\tau_i)\right)=\gamma_{0j}+\eta A_{ij}+\bgamma_{X}^T\bX_{ijk}+\tau_i, 
\end{eqnarray}
where ${\rm logit}(x)=\log \bigr(\frac{x}{1-x}\bigr)$ is the link function for $h$. According to the definitions of NIE and NDE in (\ref{ave_def}), for calendar time index $j$, we obtain
\begin{eqnarray*}
	\text{NIE}(j|\bx)&=&\beta_{M}\bigr[\kappa(1,j|\bx)-\kappa(0,j|\bx)\bigr],\\
	\text{NDE}(j|\bx)&=&\theta,
\end{eqnarray*}
where the details are shown in Web Appendix A2, and the quantity that defines the NIE during period $j$ is given by the single integral over the random-effects distribution in the mediator model, 
\begin{equation}\label{eq:kappa}
	\kappa(a,j|\bx)=\int \frac{\exp(\gamma_{0j}+\eta a+\bgamma_{X}^T\bx+\tau)}{1+\exp(\gamma_{0j}+\eta a+\bgamma_{X}^T\bx+\tau)} \frac{1}{\sqrt{2}\sigma_{\tau}}\exp\bigg(-\frac{\tau^2}{2\sigma_{\tau}^2}\bigg)d\tau,~~~a\in\{0,1\}.
\end{equation}
Under this scenario, the quantity $\text{NIE}(j|\bx)$ depends on calendar time $j$ and covariates, but $\text{NDE}(j|\bx)$ are free of both $j$ and covariates. Therefore, the overall summary mediation effect measures based on uniform period averages are explicitly given by
\begin{eqnarray*}
	\text{NIE}(\bx)&=&\frac{1}{J}\sum_{j=1}^J\text{NIE}(j|\bx)=\frac{1}{J}\sum_{j=1}^J \beta_{M}\bigr[\kappa(1,j|\bx)-\kappa(0,j|\bx)\bigr],~~~
	\text{NDE}(\bx)=\frac{1}{J}\sum_{j=1}^J\text{NDE}(j|\bx)=\theta,\\
	\text{MP}(\bx)&=&\frac{\sum_{j=1}^J \bigr[\kappa(1,j|\bx)-\kappa(0,j|\bx)\bigr]}{\sum_{j=1}^J\bigg[\kappa(1,j|\bx)-\kappa(0,j|\bx)\bigg]+J\beta_{M}^{-1}\theta}.
\end{eqnarray*}

To obtain NIE, NDE and MP, one can first fit models (\ref{cont_Y}) and (\ref{bina_M}) in standard software (e.g. \texttt{nlme}, \texttt{lme4} or \texttt{glmmTMB} R packages) to obtain corresponding coefficient estimators and hence NDE. The overall summary $\text{NIE}(\bx)$, involves a one-dimensional logistic-normal integral $\kappa(a,j|\bx)$ ---that does not have a closed-form solution. We consider two methods to approximate this logistic-normal integral---the standard Gauss-Hermite Quadrature (GHQ) \citep{Liu1994}, and the second-order Taylor approximation (STA)---that was used in approximate maximum likelihood estimation to correct for measurement error in covariates for logistic regression by \citet{Cao2021}. For example, when using the STA technique to approximate the $\kappa(a,j|\bx)$, we have
\begin{eqnarray*}
	\kappa(a,j|\bx)\approx m(a,j|\bx)+[m(a,j|\bx)-3m^2(a,j|\bx)+2m^3(a,j|\bx)]\frac{1}{2}\sigma_{\tau}^2,
\end{eqnarray*}
where $m(a,j|\bx) = \{\exp(\gamma_{0j}+\eta a+\bgamma_{X}^T\bx)\}/\{1+\exp(\gamma_{0j}+\eta a+\bgamma_{X}^T\bx)\}$ is the logistic likelihood for a given data point without the random effect. Compared to the standard GHQ approach, the STA approach tends to be much more computationally efficient, and we will examine both approaches in the ensuing simulations. 

\subsection{Data type 3---A binary outcome and a continuous mediator ($Y_bM_c$)}\label{sec:bc_hhm}
Under this scenario, the mediator model remains the same as model (\ref{cont_M}), but we assume the binary outcome to follow a logistic generalized linear mixed model as:
\begin{eqnarray}\label{bina_Y}
	{\rm logit}\left(P(Y_{ijk}=1|A_{ij},M_{ijk},\bX_{ijk},\alpha_i)\right)=\beta_{0j}+\theta A_{ij}+\beta_{M}M_{ijk}+\bbeta_X^T\bX_{ijk}+\alpha_i.
\end{eqnarray}
In Web Appendix A3, we derive exact expressions for the mediation measures under the conditions that $M_{ijk}|\{A_{ij},\bX_{ijk},\tau_i\}$ in model (\ref{cont_M}) follows a normal distribution. To present the mediation effect measures, for $a,a^*\in \{0,1\}$ and conditional on covariate values $\bx$, we define a calendar time-specific double integral
\begin{eqnarray}
	\mu(a,a^*,j|\bx)&=&P\left(Y_{ijk}(a,M_{ijk}(a^*))=1|\bX_{ijk}=\bx\right)\nonumber \\
	&=&\int\bigg[\int\frac{\exp(\beta_{0j}+\theta a+\beta_{M}m+\bbeta_X^T\bx+\alpha)}{1+\exp(\beta_{0j}+\theta a+\beta_{M}m+\beta_X^T\bx+\alpha)} \frac{1}{\sqrt{2\pi(\sigma_{\tau}^2+\sigma_e^2)}}\nonumber \\ &&\times \exp\bigg(-\frac{(m-\gamma_{0j}-\eta a^*-\bgamma_{X}^T\bx)^2}{2(\sigma_{\tau}^2+\sigma_e^2)}\bigg)dm\bigg]\cdot\frac{1}{\sqrt{2}\sigma_{\alpha}}\exp\bigg(-\frac{\alpha^2}{2\sigma_{\alpha}^2}\bigg)d\alpha\nonumber \\
	&=&\int E_m\left\{q(m|\bx,\alpha)\right\}\times\frac{1}{\sqrt{2}\sigma_{\alpha}}\exp\bigg(-\frac{\alpha^2}{2\sigma_{\alpha}^2}\bigg)d\alpha,\label{eq:d-int}
\end{eqnarray}
where $q(m|\bx,\alpha)=\frac{\exp(\beta_{0j}+\theta a+\beta_{M}m+\bbeta_X^T\bx+\alpha)}{1+\exp(\beta_{0j}+\theta a+\beta_{M}m+\bbeta_X^T\bx+\alpha)}$, and $E_m\{\bullet\}$ is defined with respect to the normal distribution of mediator marginalized over the mediator random effect, that is, a normal density with mean $\overline{m}=\gamma_{0j}+\eta a^*+\bgamma_{X}^T\bx$ and variance $\sigma_{\tau}^2+\sigma_e^2$. Then the period-specific NIE and NDE are given by
\begin{eqnarray*}
	\text{NIE}(j|\bx)&=&\log\bigg[\frac{\mu(1,1,j|\bx)}{1-\mu(1,1,j|\bx)}\bigg]-\log\bigg[\frac{\mu(1,0,j|\bx)}{1-\mu(1,0,j|\bx)}\bigg],\\
	\text{NDE}(j|\bx)&=&\log\bigg[\frac{\mu(1,0,j|\bx)}{1-\mu(1,0,j|\bx)}\bigg]-\log\bigg[\frac{\mu(0,0,j|\bx)}{1-\mu(0,0,j|\bx)}\bigg].
\end{eqnarray*}
The detailed derivations of $\text{NIE}(j|\bx)$, $\text{NDE}(j|\bx)$ are shown in Web Appendix A3. The overall summary mediation effect measures can then be obtained through $\text{NIE}(\bx)=\frac{1}{J}\sum_{j=1}^J \text{NIE}(j|\bx)$, $\text{NDE}(\bx)=\frac{1}{J}\sum_{j=1}^J \text{NDE}(j|\bx)$, $\text{MP}(\bx)=\text{NIE}(\bx)/\{\text{NIE}(\bx)+\text{NDE}(\bx)\}$, and once the estimates for all model coefficients are obtained, the plug-in estimates for all mediation effect measures can be generated.

A practical challenge to obtain the above mediation effect estimates is that one would need to address $\mu(a,a^*,j|\bx)$ as a double integral, which is computationally burdensome to numerically approximate through the GHQ approach. To circumvent the need for complex numerical approximation, we propose using the double-STA method, that is, using twice STA method to approximate the double integral $\mu(a,a^*,j|\bx)$, which generalizes the STA method used in Section \ref{sec:cb_hhm}. For brevity, we provide the detailed derivation of the double-STA method in Web Appendix B. 

\subsection{Data type 4---A binary outcome and a binary mediator ($Y_bM_b$)}\label{sec:bb_hhm}
When the outcome model and the mediator model are specified in (\ref{bina_Y}) and (\ref{bina_M}) as a pair of logistic generalized linear mixed models (such that both $g$ and $h$ are logistic link functions), respectively, we show in Web Appendix A4 that the period-specific mediation effect measures are given by
\begin{eqnarray*}
	\text{NIE}(j|\bx)&=&\log\bigg\{\frac{\lambda(1,0,j|\bx)[1-\kappa(1,j|\bx)]+\lambda(1,1,j|\bx)\kappa(1,j|\bx)}{1-[\lambda(1,0,j|\bx)[1-\kappa(1,j|\bx)]+\lambda(1,1,j|\bx)\kappa(1,j|\bx)]}\bigg\}\nonumber \\
	&&-\log\bigg\{\frac{\lambda(1,0,j|\bx)[1-\kappa(0,j|\bx)]+\lambda(1,1,j|\bx)\kappa(0,j|\bx)}{1-[\lambda(1,0,j|\bx)[1-\kappa(0,j|\bx)]+\lambda(1,1,j|\bx)\kappa(0,j|\bx)]}\bigg\},\\
	\text{NDE}(j|\bx)&=&\log \bigg\{\frac{\lambda(1,0,j|\bx)[1-\kappa(0,j|\bx)]+\lambda(1,1,j|\bx)\kappa(0,j|\bx)}{1-[\lambda(1,0,j|\bx)[1-\kappa(0,j|\bx)]+\lambda(1,1,j|\bx)\kappa(0,j|\bx)]}\bigg\}\nonumber \\
	&&-\log\bigg\{\frac{\lambda(0,0,j|\bx)[1-\kappa(0,j|\bx)]+\lambda(0,1,j|\bx)\kappa(0,j|\bx)}{1-[\lambda(0,0,j|\bx)[1-\kappa(0,j|\bx)]+\lambda(0,1,j|\bx)\kappa(0,j|\bx)]}\bigg\},
\end{eqnarray*}
where $\kappa(a,j|\bx)$ is defined in equation \eqref{eq:kappa}, and 
\begin{eqnarray}\label{eq:lambda}
	\lambda(a,a^*,j|\bx)=\int\frac{\exp\left(\beta_{0j}+\theta a+\beta_Ma^*+\bbeta_X^T\bx+\alpha\right)}{1+\exp\left(\beta_{0j}+\theta a+\beta_Ma^*+\bbeta_X^T\bx+\alpha\right)}\frac{1}{\sqrt{2}\sigma_{\alpha}}\exp\bigg(-\frac{\alpha^2}{2\sigma_{\alpha}^2}\bigg)d\alpha,\qquad \qquad a,a^*\in \{0,1\}.
\end{eqnarray}
Then, the period-specific $\text{MP}(j|\bx)$ can be directly obtained by ${\text{NIE}(j|\bx)}/[{\text{NIE}(j|\bx)+\text{NDE}(j|\bx)}]$, and all the summary across periods can be obtained via the general expressions in \eqref{ave_def_total}. 
To implement these expressions based on model parameter estimates, we notice that both $\lambda(a,a^*,j|\bx)$ and $\kappa(a,j|\bx)$ are logistic-normal single integrals, and therefore can be estimated via either the GHQ or STA methods. The details on approximating the single integral with STA follow those in Section \ref{sec:cb_hhm}.

For ease of reference, we summarize the period-specific $\text{NIE}(j|\bx)$ and $\text{NDE}(j|\bx)$ expressions under each data type in the first part of Table \ref{tb:summary1}.

\begin{center}
	[Table \ref{tb:summary1} about here.]
\end{center}

\subsection{Variance estimation via cluster jackknifing}\label{sec:var_jackknife}
Except for the $Y_cM_c$ data type, the expressions of NIE and NDE are generally complicated; although derivations for an asymptotic variance formula may be possible, they are in general cumbersome to execute given that some mediation effect measures are defined based a single or double integral. To provide a general recipe for inference, we consider the cluster jackknife estimator originally developed by \citet{Tukey1958}. Accounting for correlation of outcomes and mediators within the same cluster, the jackknife variance estimator of an estimator for parameter $\bxi\in\{\rm NIE,NDE,TE,MP\}$ can be computed as 
\begin{eqnarray}\label{jackknife}
	{\rm \widehat{Var}}(\widehat{\bxi})=\frac{I-1}{I}\sum_{i=1}^I(\widehat{\bxi}_{-i}-\overline{\bxi})^2,
\end{eqnarray}
where $\overline{\bxi}=\frac{1}{I}\sum_{i=1}^I\widehat{\bxi}_{-i}$, and $\widehat{\bxi}_{-i}$ is the delete-one-cluster estimator obtained by removing the observations in cluster $i$. Based on the variance estimator, the $100(1-\alpha)\%$ confidence interval can be constructed from  
$\widehat{\bxi}\pm t_{I-1}^{1-\alpha/2}\sqrt{\rm\widehat{Var}(\widehat{\bxi})}$, where $t_{I-1}^{1-\alpha/2}$ is the $1-\alpha/2$ quantile of the $t$ distribution with degrees of freedom $I-1$.

\section{Mediation analysis under an exposure-time dependent treatment effect structure}\label{sec:td_te}
\subsection{Mediation effect measures and assumptions}
In this Section, we generalize the proposed mediation analysis methods to further accommodate an exposure-time dependent treatment effect structure. Such a treatment effect structure may be plausible if there exists a delayed or learning effect over duration of the treatment implementation, a weakening of effect over time due to fatigue, among others. While the analysis of SW-CRTs has been conventionally assumed an instantaneous and constant treatment effect structure, \citet{Hughes2015} first discussed the possibility of modeling treatment effect as a function of number of periods elapsed since treatment adoption; also see Section 3.3 of \citet{Li2021} for a review and visualization of possible treatment effect structures. \citet{Kenny2022} pointed out an important observation that ignoring the time dependent treatment structure when it exists can even lead to bias of the treatment effect estimate toward the opposite direction, and suggested careful inclusion of the time dependent treatment effect parameters in the linear mixed model. \citet{Maleyeff2023} provided a further comparison between the fixed and random time dependent treatment effect models in SW-CRTs, and suggested a new test for exposure-time treatment effect heterogeneity. \citet{wang2024achieve} provided a rigorous causal inference framework for analyzing SW-CRTs under a constant and time-dependent treatment effect structure, and found that correctly specifying this treatment effect structure is more essential for model-robust inference than other model components. 

Building on these prior literature, we consider the following paired generalized linear mixed models (\ref{td_models}) allowing for exposure-time treatment effect heterogeneity for both the outcomes and the mediators:
\begin{eqnarray}\label{td_models}
	g\left(E(Y_{ijk}|A_{ij},M_{ijk},\bX_{ijk},\alpha_i)\right)&=&\beta_{0j}+\theta_{E_{ij}}A_{ij}+\beta_{M}M_{ijk}+\bbeta_X^T\bX_{ijk}+\alpha_i,\nonumber\\
	h\left(E(M_{ijk}|A_{ij},\bX_{ijk},\tau_i)\right)&=&\gamma_{0j}+\eta_{E_{ij}} A_{ij}+\bgamma_X^T\bX_{ijk}+\tau_i,
\end{eqnarray}
where $\theta_{E_{ij}}=\theta(E_{ij})$ and $\eta_{E_{ij}}=\eta(E_{ij})$ with $E_{ij}\in\{0,1,\ldots,E\}$, and $\theta(0)=\eta(0)=0$ by definition. Therefore, $\theta_{E_{ij}}$ and $\eta_{E_{ij}}$ are distinct values for each level of exposure time to accommodate an arbitrary pattern of the treatment effect curve. 
To define the mediation measures under this context, since the treatment effect parameters are exposure-time dependent, we define $Y_{ijk}(e)$ and $M_{ijk}(e)$ as the potential outcome and mediator for each individual when cluster $i$ has been treated for $E_{ij}=e$ periods already. Notice that $Y_{ijk}(e)$ and $M_{ijk}(e)$ are only defined for $0\leq e\leq j-1$, but no larger values of $e$ due to the feature of the stepped wedge design (assuming a single baseline period as is common in practice).  
Similarly, we denote $Y_{ijk}(e,m)$ as the potential outcome $Y_{ijk}$ when setting $M_{ijk}=m$ under exposure time $E_{ij}=e$. Then, extending the notation in Section \ref{sec:assumption1}, if the treatment level is changed from control to an intervention for $e$ periods, the NIE and NDE during calendar time $j$ can be defined on the $g$-function scale:
\begin{eqnarray}\label{ave_def_td}
	\text{NIE}(j,e|\bx)&=&g\left(E[Y_{ijk}(e,M_{ijk}(e))|\bX_{ijk}=\bx]\right)-g\left(E[Y_{ijk}(e,M_{ijk}(0))|\bX_{ijk}=\bx]\right),\nonumber \\
	\text{NDE}(j,e|\bx)&=&g\left(E[Y_{ijk}(e,M_{ijk}(0))|\bX_{ijk}=\bx]\right)-g\left(E[Y_{ijk}(0,M_{ijk}(0))|\bX_{ijk}=\bx]\right),
\end{eqnarray}
for $1\leq e\leq j-1$ and $2 \leq j \leq J$.
Therefore, the MP during period $j$ by contrasting exposure time $e$ and control is given by the ratio, $\text{MP}(j,e|\bx)=\text{NIE}(j,e|\bx)/\text{TE}(j,e|\bx)=\text{NIE}(j,e|\bx)/\{\text{NIE}(j,e|\bx)+\text{NDE}(j,e|\bx)\}$. 

We next define summary mediation effect measures. 
Different from the instantaneous and constant treatment effect structure, the $\text{NIE}(j,e|\bx)$ and $\text{NDE}(j,e|\bx)$ depend on both the calendar time and exposure time, and we can summarize the effects across these two time dimensions. We first consider summarizing the effect over calendar time, for a fixed 
exposure time $e \in \{1,2,\dots,J-1\}$. Note that according to panel (a) of Figure \ref{fig:scheme}, eligible calendar periods that correspond to exposure time $e$ are those such that $j\geq e+1$, and the exposure-time specific NIE and NDE are given by 
\begin{eqnarray}\label{def_td}
	\text{NIE}(e|\bx)&=&\sum_{j=e+1}^J \omega(j,e)\text{NIE}(j,e|\bx),~~~
	\text{NDE}(e|\bx)=\sum_{j=e+1}^J\omega(j,e)\text{NDE}(j,e|\bx),\nonumber \\
	\text{MP}(e|\bx)&=&\sum_{j=e+1}^J \left\{\frac{\text{NIE}(j,e|\bx)+\text{NDE}(j,e|\bx)}{\sum_{l=e+1}^J\text{NIE}(l,e|\bx)+\text{NDE}(l,e|\bx)}\right\}\text{MP}(j,e|\bx).
\end{eqnarray} 
We primarily focus on the uniform weight $\omega(j,e)=1/(J-e)$ for simplicity and interpretability although there may exist different possibilities in specifying the weights $\omega(j,e)$ to flexibly determine the relative contribution from each period to the exposure time. Next, to obtain a single summary over levels of exposure time units, the summary mediation effect measures over both calendar and exposure time are given by 
\begin{eqnarray}\label{def_td_total}
	\text{NIE}(\bx)&=&\frac{1}{E}\sum_{e=1}^E\text{NIE}(e|\bx)
	,~~~
	\text{NDE}(\bx)=\frac{1}{E}\sum_{e=1}^E\text{NDE}(e|\bx)
	,\nonumber \\
	\text{MP}(\bx)&=&\sum_{e=1}^{E} \left\{\frac{\text{NIE}(e|\bx)+\text{NDE}(e|\bx)}{\sum_{r=1}^{E}\text{NIE}(r|\bx)+\text{NDE}(r|\bx)}\right\}\text{MP}(e|\bx).
\end{eqnarray}
The features and interpretations of these mediation effect measures are similar to those developed in Section \ref{sec:assumption1}, except that the average is defined with respect to both the calendar time and exposure time, as a result of exposure-time treatment effect heterogeneity.

To identify NIE and NDE defined in (\ref{ave_def_td}), we need to extend the structural assumptions discussed in Section \ref{sec:assumption1} by now using the modified potential outcomes notation as a function of the exposure time. 
The assumptions include (A.5) when we set $E_{ij}=e\in\{0,1,\ldots,E\}$, the observed outcome and mediator become $Y_{ijk}=Y_{ijk}(e)$ and $M_{ijk}=M_{ijk}(e)$; furthermore, the observed outcome $Y_{ijk}=Y_{ijk}(e,m)$ when we set $E_{ij}=e$ and $M_{ijk}=m$; (A.6) the potential outcome under $E_{ij}=e$ is equal to the potential outcome when $E_{ij}=e$ and $M_{ijk}$ is set to its natural value under $E_{ij}=e$, that is, $Y_{ijk}(e)=Y_{ijk}(e,M_{ijk}(e))$; (A.7) (i) $Y_{ijk}(e,m) \perp E_{ij}|\bX_{ijk}$; (ii) $Y_{ijk}(e,m) \perp M_{ijk}|(\bX_{ijk},E_{ij})$; (iii) $M_{ijk}(e) \perp E_{ij}|\bX_{ijk}$; (A.8) For all $e$, $e^*$ and $m$, $Y_{ijk}(e,m)\perp M_{ijk}(e^*)|\bX_{ijk}$. The roles of assumptions (A.5)-(A.8) are the same as (A.1)-(A.4), except that they are now stated based on potential outcomes defined as a function of treatment duration. To elaborate, Assumption (A.5) is the causal consistency assumption stated for each level of exposure time. Assumption (A.6) is the composition assumption. Assumption (A.7) is the conditional independence assumption, and parts (i),(ii) and (iii) assume that the treatment-outcome, mediator-outcome and treatment-mediator relationships are unconfounded given the measured covariates $\bX_{ijk}$. Importantly, part (i) and part (iii) under (A.7) also hold automatically in SW-CRTs because the clusters are randomized to distinct treatment sequences. Therefore, within each period $j$, the exposure time $E_{ij}$ is independent of the potential mediator and potential outcome (since there is a one-to-one mapping between $E_{ij}$ and each distinct treatment sequence). Similar to Section \ref{sec:assumption1}, we only require the collection of measured covariates to deconfound the mediator-outcome relationship. Finally, (A.8) assumes that none of the confounders in the mediator-outcome relationship can be affected by the exposure time $E_{ij}$.
Under these structural assumptions and the paired generalized linear mixed models \eqref{td_models}, we next derive explicit expressions of the mediation effect measures, by considering all four data types.

\subsection{Data type 1---A continuous outcome and a continuous mediator ($Y_cM_c$)}\label{sec:cc_etm}
We set link functions $g$ and $h$ in (\ref{td_models}) to be the identity functions when both the outcome and mediator are continuous variables, and write the outcome model and the mediator model as the following pair of linear mixed models
\begin{eqnarray}
	Y_{ijk}&=&\beta_{0j}+\theta_{E_{ij}}A_{ij}+\beta_{M}M_{ijk}+\bbeta_X^T\bX_{ijk}+\alpha_i+\epsilon_{ijk},\label{cont_Y_etm}\\
	M_{ijk}&=&\gamma_{0j}+\eta_{E_{ij}}A_{ij}+\bgamma_{X}^T\bX_{ijk}+\tau_i+e_{ijk}, \label{cont_M_etm}
\end{eqnarray} 
where $\epsilon_{ijk}\sim N(0,\sigma^2_{\epsilon})$ and $e_{ijk} \sim N(0,\sigma^2_{e})$ are independent residual error terms in the outcome model and mediator model, respectively. By the definitions of mediation effect measures in (\ref{ave_def_td}), we obtain from models \eqref{cont_Y_etm} and \eqref{cont_M_etm} that
\begin{eqnarray*}
	\text{NIE}(j,e|\bx)
	&=&\beta_{M}\eta_e ,\\
	\text{NDE}(j,e|\bx)&=&\theta_e,  \qquad  1\leq e\leq j-1 ~~\text{and}~~2 \leq j \leq J.
\end{eqnarray*}
Since both $\text{NIE}(j,e|\bx)$ and $\text{NDE}(j,e|\bx)$ are free of calendar time $j$ and covariates, 
the exposure-time specific $\text{NIE}(e|\bx)$ and $\text{NDE}(e|\bx)$ are equal to $\beta_{M}\eta_e$ and $\theta_e$, respectively, and the $\text{MP}(e|\bx)={\beta_{M}\eta_e}/{(\beta_{M}\eta_e+\theta_e)}$. Furthermore, according to definitions in (\ref{def_td_total}), the overall summary
mediation effect measures across each exposure times are given 
\begin{eqnarray*}
	\text{NIE}(\bx)&=&\frac{1}{J-1}\sum_{e=1}^{J-1}\beta_{M}\eta_e,~~~\text{NDE}(\bx)=\frac{1}{J-1}\sum_{e=1}^{J-1}\theta_e,~~~\text{MP}(\bx)=\frac{\beta_{M}\sum_{e=1}^{J-1}\eta_e}{\sum_{e=1}^{J-1}(\beta_{M}\eta_e+\theta_e)}.
\end{eqnarray*}
Expressions of each exposure-time specific NIE and NDE are similar to those under an instantaneous and constant treatment effect structure, carrying a form in the product of coefficients. To obtain exposure-time dependent NIE, NDE and MP and corresponding mediation effect measures $\text{NIE}(\bx)$, $\text{NDE}(\bx)$ and $\text{MP}(\bx)$, one can fit models \eqref{cont_Y_etm} and \eqref{cont_M_etm} using maximum likelihood or restricted maximum likelihood approaches (e.g. with the \texttt{nlme} or \texttt{lme4} R package), and plug in the respective coefficient estimators. 

\subsection{Data type 2---A continuous outcome and a binary mediator ($Y_cM_b$)}\label{sec:cb_etm}
When the outcome model of $Y_{ijk}$ remains the same as model (\ref{cont_Y_etm}) and the mediator is binary, we apply the following logistic generalized linear mixed model to model the mediator, that is
\begin{eqnarray}\label{bina_M_etm}
	{\rm logit}\left(P(M_{ijk}=1|A_{ij},\bX_{ijk},\tau_i)\right)=\gamma_{0j}+\eta_{E_{ij}} A_{ij}+\bgamma_{X}^T\bX_{ijk}+\tau_i. 
\end{eqnarray}
Then, according to the definitions of (\ref{ave_def_td}), under eligible calendar time index $j$ at exposure time $e$, we obtain
\begin{eqnarray*}
	\text{NIE}(j,e|\bx)&=&\beta_{M}\bigr[\kappa_e(1,j|\bx)-\kappa_e(0,j|\bx)\bigr],\\
	\text{NDE}(j,e|\bx)&=&\theta_e, \qquad 1\leq e\leq j-1 ~~\text{and}~~2 \leq j \leq J,
\end{eqnarray*}
where $\kappa_e(a,j|\bx)$ is given by the single integral over the random-effects distribution in the mediator model, 
\begin{equation}\label{eq:kappa_etm}
	\kappa_e(a,j|\bx)=\int \frac{\exp(\gamma_{0j}+\eta_e a+\bgamma_{X}^T\bx+\tau)}{1+\exp(\gamma_{0j}+\eta_e a+\bgamma_{X}^T\bx+\tau)} \frac{1}{\sqrt{2}\sigma_{\tau}}\exp\bigg(-\frac{\tau^2}{2\sigma_{\tau}^2}\bigg)d\tau,~~~a\in\{0,1\}.
\end{equation}
Under this data type, the mediation effect measure $\text{NIE}(j,e|\bx)$ depends on calendar time $j$ and covariates, but $\text{NDE}(j,e|\bx)$ are free of both $j$ and covariates. Therefore, by (\ref{def_td}), the exposure-time specific mediation effect measures are 
\begin{eqnarray*}
	\text{NIE}(e|\bx)&=&\frac{1}{J-e}\sum_{j=e+1}^J \beta_{M}\bigr[\kappa_e(1,j|\bx)-\kappa_e(0,j|\bx)\bigr],~~~
	\text{NDE}(e|\bx)=\frac{1}{J-e}\sum_{j=e+1}^J\theta_e=\theta_e,
	\nonumber \\
	\text{MP}(e|\bx)&=&\frac{\sum_{j=e+1}^J \beta_{M}\bigr[\kappa_e(1,j|\bx)-\kappa_e(0,j|\bx)\bigr]}{\sum_{j=e+1}^J \bigr\{\beta_{M}\bigr[\kappa_e(1,j|\bx)-\kappa_e(0,j|\bx)\bigr]+\theta_e\bigr\}}, \qquad  1\leq e\leq J-1.
\end{eqnarray*}

To obtain NIE, NDE and MP, one can first fit models (\ref{cont_Y_etm}) and (\ref{bina_M_etm}) in standard software (e.g. \texttt{nlme}, \texttt{lme4} or \texttt{glmmTMB} R packages) to obtain corresponding coefficient estimators and hence $\text{NDE}(j,e|\bx)$. The calendar-exposure-time specific NIE, $\text{NIE}(j,e|\bx)$, involves a one-dimensional logistic-normal integral $\kappa_e(a,j|\bx)~(a\in\{0,1\})$. Although this integral $\kappa_e(a,j|\bx)$ does not have a closed-form solution, we apply the GHQ and the STA techniques described in Section \ref{sec:const_te} to approximate it. After obtaining all exposure-time specific mediation effect measures $\text{NIE}(e|\bx)$ and $\text{NDE}(e|\bx)$, the overall summary mediation effect measures based on uniform exposure-time averages (i.e., $\text{NIE}(\bx)$, $\text{NDE}(\bx)$ and $\text{MP}(\bx)$) can be obtained directly through (\ref{def_td_total}).

\subsection{Data type 3---A binary outcome and a continuous mediator ($Y_bM_c$)}\label{sec:bc_etm}
Under this scenario, the mediator model is the same as that in (\ref{cont_M_etm}). For the binary outcome, we apply a logistic generalized linear mixed model to model that, \begin{eqnarray}\label{bina_Y_etm}
	{\rm logit}\left(P(Y_{ijk}=1|A_{ij},M_{ijk},\bX_{ijk},\alpha_i)\right)=\beta_{0j}+\theta_{E_{ij}} A_{ij}+\beta_{M}M_{ijk}+\bbeta_X^T\bX_{ijk}+\alpha_i.
\end{eqnarray} 
By using the same technique in Web Appendix A3 and Section \ref{sec:bc_hhm}, the NIE and NDE during calendar time $j$ at exposure time $e$ defined in (\ref{ave_def_td}) are given by 
\begin{eqnarray*}
	\text{NIE}(j,e|\bx)&=&\log\bigg[\frac{\mu_e(1,1,j|\bx)}{1-\mu_e(1,1,j|\bx)}\bigg]-\log\bigg[\frac{\mu_e(1,0,j|\bx)}{1-\mu_e(1,0,j|\bx)}\bigg],\\
	\text{NDE}(j,e|\bx)&=&\log\bigg[\frac{\mu_e(1,0,j|\bx)}{1-\mu_e(1,0,j|\bx)}\bigg]-\log\bigg[\frac{\mu_e(0,0,j|\bx)}{1-\mu_e(0,0,j|\bx)}\bigg],\qquad 1\leq e\leq j-1 ~~\text{and}~~2 \leq j \leq J,
\end{eqnarray*}
where 
\begin{eqnarray} \label{eq:mu_etm}
	\mu_e(a,a^*,j)&=&P(Y_{ijk}(e,M_{ijk}(e^*))=1|\bX_{ijk}=\bx)\nonumber \\
	&=&\int \bigg[\int\frac{\exp(\beta_{0j}+\theta_e a+\beta_{M}m+\bbeta_X^T\bx+\alpha)}{1+\exp(\beta_{0j}+\theta_e a+\beta_{M}m+\bbeta_X^T\bx+\alpha)} \frac{1}{\sqrt{2\pi(\sigma_{\tau}^2+\sigma_ e^2)}}\\
	&&\times\exp\bigg(-\frac{(m-\gamma_{0j}-\eta_e a^*-\bgamma_{X}^T\bx)^2}{2(\sigma_{\tau}^2+\sigma_e^2)}\bigg)dm\bigg] 
	\cdot\frac{1}{\sqrt{2}\sigma_{\alpha}}\exp\bigg(-\frac{\alpha^2}{2\sigma_{\alpha}^2}\bigg)d\alpha, \qquad a,a^* \in \{0,1\}, \nonumber
\end{eqnarray}
is a calendar time-specific double integral. The exposure-time specific mediation effect measures $\text{NIE}(e|\bx)$, $\text{NDE}(e|\bx)$ and $\text{MP}(e|\bx)$ can then be obtained based on (\ref{def_td}). After that, the overall summary mediation effect measures based on uniform exposure-time averages, i.e., $\text{NIE}(\bx)$, $\text{NDE}(\bx)$ and $\text{MP}(\bx)$, can be obtained through (\ref{def_td_total}). 

In practice, to obtain an estimator for the above mediation effect measure, we need to address the double integral $\mu_e(a,a^*,j|\bx)$ and also apply the double-STA technique to circumvent the need for complex numerical approximation of $\mu_e(a,a^*,j|\bx)$; details of the double-STA technique can be found in Section \ref{sec:bc_hhm}.

\subsection{Data type 4---A binary outcome and a binary mediator ($Y_bM_b$)}\label{sec:bb_etm}
When the outcome model and the mediator model are correctly specified in (\ref{bina_Y_etm}) and (\ref{bina_M_etm}) as a pair of logistic generalized linear mixed models, respectively, both $g$ and $h$ are logistic link functions under this scenario. Then, by using the same technique in Web Appendix A4 for period-specific $\text{NIE}(j|\bx)$ and $\text{NDE}(j|\bx)$ under an instantaneous and constant treatment effect structure, we obtain the NIE and NDE during calendar time $j$ at exposure time $e$ as 
\begin{eqnarray*}
	\text{NIE}(j,e|\bx)
	&=&\log\bigg\{\frac{\lambda_e(1,0,j|\bx)[1-\kappa_e(1,j|\bx)]+\lambda_e(1,1,j|\bx)\kappa_e(1,j|\bx)}{1-[\lambda_e(1,0,j|\bx)[1-\kappa_e(1,j|\bx)]+\lambda_e(1,1,j|\bx)\kappa_e(1,j|\bx)]}\bigg\}\\
	&&-\log\bigg\{\frac{\lambda_e(1,0,j|\bx)[1-\kappa_e(0,j|\bx)]+\lambda_e(1,1,j|\bx)\kappa_e(0,j|\bx)
	}{1-[\lambda_e(1,0,j|\bx)[1-\kappa_e(0,j|\bx)]+\lambda_e(1,1,j|\bx)\kappa_e(0,j|\bx)]}\bigg\},\\
	\text{NDE}(j,e|\bx)
	&=&\log\bigg\{\frac{\lambda_e(1,0,j|\bx)[1-\kappa_e(0,j|\bx)]+\lambda_e(1,1,j|\bx)\kappa_e(0,j|\bx)}{1-[\lambda_e(1,0,j|\bx)[1-\kappa_e(0,j|\bx)]+\lambda_e(1,1,j|\bx)\kappa_e(0,j|\bx)]}\bigg\}\\
	&&-\log\bigg\{\frac{ \lambda_e(0,0,j|\bx)[1-\kappa_e(0,j|\bx)]+\lambda_e(0,1,j|\bx)\kappa_e(0,j|\bx)}{1-[\lambda_e(0,0,j|\bx)[1-\kappa_e(0,j|\bx)]+\lambda_e(0,1,j|\bx)\kappa_e(0,j|\bx)]}\bigg\},
\end{eqnarray*}
where $1\leq e\leq j-1$ and $2 \leq j \leq J$, $\kappa_e(a,j|\bx)(a\in \{0,1\})$ is defined in (\ref{eq:kappa_etm}), and 
\begin{eqnarray}\label{eq:lambda_etm}
	\lambda_e(a,a^*,j|\bx)=\int\frac{\exp\left(\beta_{0j}+\theta_e a+\beta_Ma^*+\bbeta_X^T\bx+\alpha\right)}{1+\exp\left(\beta_{0j}+\theta_e a+\beta_Ma^*+\bbeta_X^T\bx+\alpha\right)}\frac{1}{\sqrt{2}\sigma_{\alpha}}\exp\bigg(-\frac{\alpha^2}{2\sigma_{\alpha}^2}\bigg)d\alpha,\qquad \qquad a,a^*\in \{0,1\}.
\end{eqnarray}
Then, the exposure-time specific mediation effect measures can be directly obtained from (\ref{def_td}) with $\omega(i,e)=\frac{1}{J-e}$, and the overall summary mediation effect measures $\text{NIE}(\bx)$, $\text{NDE}(\bx)$ and $\text{MP}(\bx)$, can be obtained through (\ref{def_td_total}). Similar to Section \ref{sec:bb_hhm}, we estimate the two logistic-normal single integrals $\lambda_e(a,a^*,j|\bx)$ and $\kappa_e(a,j|\bx)$ by either the GHQ or STA methods. 

For ease of reference, we also summarize the calendar-exposure-time specific mediation measures $\text{NIE}(j,e|\bx)$ and $\text{NDE}(j,e|\bx)$ under each data type in the second part of Table \ref{tb:summary1}. 

\subsection{Variance estimation}
As for the variance estimation, we can also use the same jackknife variance estimator defined in (\ref{jackknife}) to compute ${\rm \widehat{Var}}(\widehat{\bxi})$ and ${\rm \widehat{Var}}(\widehat{\bxi}(e|\bx))$, where $\bxi(e|\bx) \in \{\text{NIE}(e|\bx),\text{NDE}(e|\bx),\text{TE}(e|\bx),\text{MP}(e|\bx)\}$ for $e\in \{1,2,\dots,J-1\}$.

\subsection{Software implementation}

To facilitate application of our methodology, we have implemented the described mediation effect estimators under combinations of data types and treatment effect structures in the \texttt{mediateSWCRT} R package that can be accessed via \url{https://github.com/Zhiqiangcao/mediateSWCRT}. In addition, we provide a conceptual workflow that describes the steps to assess mediation via the \texttt{mediate$\_$swcrt} function in Figure \ref{fig:workflow}. 

\begin{center}
	[Figure \ref{fig:workflow} about here.]
\end{center}

\section{Simulation studies}\label{sec:simu_constant}
\subsection{Simulation design}\label{sec:simu_design_constant}
We carry out simulations to evaluate the performance of the point and interval estimators of NIE, NDE, TE and MP in finite samples under an instantaneous and constant treatment effect structure. \textcolor{black}{For brevity, we will focus on this simpler treatment effect structure to illustrate the main message. The simulation design and results under an exposure-time dependent treatment effect structure can be found in Web Appendix D, and Tables S5-S8 of Web Appendix F.} 

To represent a range of clusters likely encountered in practice, we consider 3 levels of sample sizes, that is, $I\in \{15,30,60\}$, and an equal number of $N_{ij}=20$ individuals per cluster-period. To fix ideas, we assume a total of $J=4$ periods, and a gently increasing period effects in the outcome model and mediator model such that $\beta_{01}=\gamma_{01}=0$, $\beta_{0,j+1}-\beta_{0j}=0.1\times (0.5)^{j-1}$ and $\gamma_{0,j+1}-\gamma_{0j}=0.3\times (0.5)^{j-1}$ for $j\geq 1$, respectively. For each data type and level of sample size, we consider the exposure $A$ as a binary variable. As pointed out by \cite{Li2007}, the scale and location of the independent variable will generally not affect the estimation of the mediation effect, so for simplicity, we assume there are no confounders in both the outcome and mediator models. We set TE to be 1 as true total effect and MP to be 0.25 as true mediation proportion in each data type, respectively. Then true NIE and NDE can be directly obtained as $\rm{NIE=TE\times MP}=0.25$ and $\rm{NDE=TE \times (1-MP)}=0.75$, respectively. 

For data type $Y_cM_c$ (i.e., a continuous outcome and a continuous mediator), we generate data according to models (\ref{cont_Y}) and (\ref{cont_M}) without $\bX_{ijk}$. Specifically, we assume $\sigma_{\alpha}=\sigma_{\tau}=0.334$, $\sigma_{\epsilon}=\sigma_e=1$, which results in an outcome ICC and mediator ICC of $0.1$. We set $\eta$ to 0.4, leading to an average mediator-outcome correlation to be ${\rm Corr}(A_{ij},M_{ijk})\approx 0.31$. 
We assume an equal number of clusters are randomized to each treatment sequence which determines $A_{ij}$. We then draw mediators and outcomes sequentially from
\begin{align}
	M_{ijk} &\sim N(\gamma_{0j}+\eta A_{ij}+\tau_i,\sigma_e),\nonumber 
	\\
	Y_{ijk}&\sim N(\beta_{0j}+\theta A_{ij}+\beta_MM_{ijk}+\alpha_i,\sigma_{\epsilon}).\label{eq:Y-simulate-C}
\end{align}
where $\theta={\rm (1-MP)\times TE}$ and $\beta_M={{\rm MP\times TE}}/{\eta}$. 

For data type $Y_cM_b$ (i.e., a continuous outcome and a binary mediator), the data generation process resembles that under data type $Y_cM_c$, except that a logistic linear mixed model is used to simulate the mediators. 
That is, we draw mediators from
\begin{eqnarray}
	{\rm logit}\bigr(P\bigr(M_{ijk}=1|A_{ij},\tau_i\bigr)\bigr)=\gamma_{0j}+\eta A_{ij}+\tau_i,\label{eq:M-simulate-B}
\end{eqnarray}
where the standard deviation of the random effect is $\sigma_{\tau}=0.605$ (leading to an ICC of $0.1$ on the latent response scale \citep{Eldridge2009}), and $\eta=0.4$ such that $P(M_{ijk}=1|A_{ij}=0)\approx 0.54$. 
The outcomes are still drawn from model \eqref{eq:Y-simulate-C} except that $\theta={\rm (1-MP)\times TE}$ and $\beta_M={{\rm MP\times TE}}/\left[\kappa(1,j|0)-\kappa(0,j|0)\right]$. For estimation and inference based on simulated data, we apply both GHQ and STA methods to approximate the logistic-normal integral $\kappa(a,j|0)$ for $a\in \{0,1\}$.

For data type $Y_bM_c$ (i.e., a binary outcome and a continuous mediator), the data generation process resembles that under data type $Y_cM_c$, except that a logistic linear mixed model is used to simulate the outcomes. 
That is, we use the following logistic regression to draw binary outcomes
\begin{eqnarray}\label{eq:Y-simulate-B}
	{\rm logit}\left(P(Y_{ijk}=1|A_{ij},M_{ijk},\alpha_i)\right)=\beta_{0j}+\theta A_{ij}+\beta_MM_{ijk}+\alpha_i,
\end{eqnarray}
where the standard deviation of the random effect is $\sigma_{\alpha}=0.605$ (leading to an ICC of $0.1$ on the latent response scale \citep{Eldridge2009}), and we select $\theta$ and $\beta_M$ by numerically solving the system of equations ${\rm MP\times TE}=\mathcal{NIE}(\beta_{0j},\theta,\beta_M,\gamma_{0j},\eta)$ and ${\rm (1-MP)\times TE}=\mathcal{NDE}(\beta_{0j},\theta,\beta_M,\gamma_{0j},\eta)$, where $\mathcal{NIE}(\beta_{0j},\theta,\beta_M,\gamma_{0j},\eta)$ and $\mathcal{NDE}(\beta_{0j},\theta,\beta_M,\gamma_{0j},\eta)$ refer to $\text{NIE}(j|\bx)$ and $\text{NDE}(j|\bx)$ provided in Section \ref{sec:bc_hhm} with $\mu(a,a^*,j|0)$ for $a, a^*\in \{0,1\}$. For estimation and inference based on simulated data, we apply the STA method to approximate the double logistic-normal integral $\mu(a,a^*,j|0)$.

For data type $Y_bM_b$ (i.e., a binary outcome and a binary mediator), we simulate mediators and outcomes according to model \eqref{eq:M-simulate-B} and model \eqref{eq:Y-simulate-B}, respectively. 
The values of $\theta$ and $\beta_M$ are obtained by solving the system of equations ${\rm MP\times TE}=\mathcal{NIE}(\beta_{0j},\theta,\beta_M,\gamma_{0j},\eta)$ and ${\rm (1-MP)\times TE}=\mathcal{NDE}(\beta_{0j},\theta,\beta_M,\gamma_{0j},\eta)$, where $\mathcal{NIE}(\beta_{0j},\theta,\beta_M,\gamma_{0j},\eta)$ and $\mathcal{NDE}(\beta_{0j},\theta,\beta_M,\gamma_{0j},\eta)$ refer to $\text{NIE}(j|\bx)$ and $\text{NDE}(j|\bx)$ provided in Section \ref{sec:bb_hhm} with $\lambda(a,a^*,j|0)$ and $\kappa(a,j|0)$ for $a, a^*\in \{0,1\}$. For estimation and inference based on simulated data, we then apply both GHQ and STA methods to approximate the logistic-normal integral $\kappa(a,j|0)$ and $\lambda(a,a^*,j|0)$.

For each data type, we generate 1000 data replications under each level of sample size, and fit the proposed methods for estimating mediation measures. We report the percent bias (Bias(\%))--- calculated
as the average bias relative to the true value over
1,000 replications, Monte Carlo standard deviation (MCSD), average estimated standard error based on jackknifing (AESE) and coverage probability (CP) of the 95\% confidence intervals. The simulation results for the point, variance and coverage probability of the summary NIE, NDE, TE and MP under the four data types are presented in Table \ref{tab:sim1}; simulation results for finer, period-specific mediation measures (i.e., $\text{NIE}(j|\bx)$, $\text{NDE}(j|\bx)$, $\text{TE}(j|\bx)$ and $\text{MP}(j|\bx)$) for data types $Y_cM_b$, $Y_bM_c$ and $Y_bM_b$ are provided in Tables S2-S4 of Web Appendix F, respectively.   

\subsection{Simulation results}\label{sec:simu_res}
\begin{center}
	[Table \ref{tab:sim1} about here.]
\end{center}

According to the results in Table 2, these summary mediation measures $\widehat{\bxi}=(\widehat{\rm NIE},\widehat{\rm NDE},\widehat{\rm TE},\widehat{\rm MP})^T$ exhibit minimal bias across different data types and levels of sample size. Almost all percent bias values of $\widehat{\bxi}$ are less than 1\% except for $\widehat{\rm NDE}$ and $\widehat{\rm TE}$ under data types $Y_bM_c$ and $Y_bM_b$ with a small number of clusters, $I=15$. The estimated standard errors on average are close to the Monte Carlo standard deviations of the estimator, which indicates good performance of the cluster jackknife variance estimator (\ref{jackknife}).
The coverage probability are also reasonably close to the nominal level across all simulation scenarios, and is occasionally conservative when $I=15$ under data type $Y_bM_b$. Under data types $Y_cM_b$ and $Y_bM_b$, we observe that the percent bias under the STA approach is generally smaller than that under the GHQ approach when $I=15$, but the reverse is true when $I=60$, likely due to the fact that GHQ (with increased amount of computation power) requires more data points to achieve more accurate integral approximation. However, the remaining performance metrics are almost identical when comparing STA and GHQ approaches. 
As for period-specific mediation measures, results in Tables S2-S4 of Web Appendix F show that they are all similar to their corresponding summary mediation measures under both the STA and GHQ approaches, confirming adequate performance of our estimators for $\text{NIE}(j|\bx)$, $\text{NDE}(j|\bx)$, $\text{TE}(j|\bx)$ and $\text{MP}(j|\bx)$.

\section{An Illustrative Stepped Wedge Trial Data Example}\label{sec:application}
\textcolor{black}{To provide an illustrative application of our proposed methods}, we conduct mediation analysis of a cross-sectional SW-CRT in Uganda which aims to study the impact of digital adherence technology for tuberculosis (TB) treatment supervision \citep{Crowder2020,Cattamanchi2021}.
All 18 health facilities started with routine TB treatment supervision (control condition), and in each subsequent month, 3 health facilities were randomly chosen to switch to 99DOTS-based TB treatment supervision (intervention condition), 
and 99DOTS is a low-cost digital adherence technology intervention that is designed to increase TB treatment completion. 
We focus on the outcome of TB treatment success, defined as being cured or treatment completed. In the published analysis by \citet{Cattamanchi2021}, the intention-to-treat analysis reported no statistically significant changes in treatment completion due to the 99DOTS-based intervention. Excluding patients who had early TB treatment initiation than the randomization schedule, as well as patients who did not enroll into the treatment cluster in time, the per protocol analysis reported a significantly higher treatment success rates under the 99DOTS-based intervention condition. Similarly, the per protocol analysis found that 99DOTS-based intervention 
improved the rates of completing the intensive TB treatment phase, defined as completing 60 doses. \textcolor{black}{As indicated in Figure \ref{fig:scheme}(b), mediation analysis typically relies on the presence of an observed treatment effect on both the final outcome and the intermediate outcome (potential mediator). Consequently, we investigate the question of whether the effect of intervention on the primary outcome treatment success may be mediated by the secondary outcome completed intensive phase. Notice that we frame the illustrative example on the per-protocol population due to the observed treatment effect signals on both treatment success (final outcome) and completion of the intensive phase (potential mediator). However, this should not be construed as a recommendation to conduct mediation analysis exclusively within the per-protocol population for all SW-CRTs. The per-protocol population includes 18 clusters randomized over 8 periods, and involves an implementation period. In the following analysis, we will adapt our development to accommodate an incomplete design with an implementation period. We adjust for the following potential confounders in both the outcome and mediator models: age, sex, HIV status, bacteriologically confirmed versus clinical TB diagnosis, new versus retreatment TB diagnosis. Finally, we emphasize that the purpose of the analysis is for methods illustration in a real data context. In practice, a well-designed mediation analysis should be guided by substantive knowledge or prior evidence that supports the hypothesized relationships among the treatment, mediator, and outcome variables.}

We first consider mediation analysis with a constant treatment effect structure, that is, we use (\ref{bina_Y}) and (\ref{bina_M}) as a pair of logistic generalized linear mixed models to model the outcome 
and mediator, with $\bX=\{\rm{age, sex,HIV~status,disease~dlass,retreament}\}^T$ as confounders. \textcolor{black}{To accommodate the implementation period, we set the mediators and outcomes to be missing in that period when fitting logistic generalized linear mixed models.} Then we apply $\text{NIE}(j|\bx)$ and $\text{NDE}(j|\bx)$ derived in Section \ref{sec:bb_hhm} (\textcolor{black}{invariant to the presence of an implementation period}) to obtain the four summary mediation measures. The results from both GHQ and STA methods evaluated at the median level of confounders are reported in Table 3, and 
are numerically similar. We find that the 99DOTS-based TB treatment supervision can improve TB treatment success, with odds ratios of 3.72 and 2.88 for the estimated total effect and estimated natural indirect effect, respectively; and the associated 95\% confidence intervals exclude the null. Furthermore, based on the estimated mediation proportion, about 80.6\% of the intervention effect is explained by the completion of the intensive phase. 

\begin{center}
	[Table \ref{tab:res_data} about here.]
\end{center}

As a further analysis, we investigate the exposure time-dependent mediation effects, by 
using  (\ref{bina_Y_etm}) and (\ref{bina_M_etm}) as a pair of logistic generalized linear mixed models to model outcome and mediator with the same set of confounders. \textcolor{black}{To accommodate the implementation period, the derived mediation measures in Section \ref{sec:td_te} needs to be slightly modified. That is, the maximum value of exposure time is $E=J-2=6$.
	Therefore, both subscripts $j$ and $l$ in the summation should start from $e+2$ instead of $e+1$ when estimating $\text{NIE}(e|\bx)$ $\text{NDE}(e|\bx)$ and $\text{MP}(e|\bx)$ in the definition (\ref{def_td}); also see Figure S1 in Web Appendix F for an updated design schematic}. 
The estimation results are summarized in Table 3, and suggest the NIE, NDE, TE and MP estimates appear to differ across level of exposure time. However, all mediation effects are statistically insignificant at the 0.05 level except for $\widehat{\text{NIE}}(e)$ at $e=1,2$ and $e=3$ with GHQ approach, with odds ratios of $\widehat{\text{NIE}}(1)$, $\widehat{\text{NIE}}(2)$ and $\widehat{\text{NIE}}(3)$ are 2.69, 3.41 and 4.10, respectively, indicating that the intervention effect at exposure time $e=1$, $e=2$ and $e=3$ may be partially explained by the completion of the intensive phase variable. \textcolor{black}{There are several seemingly unusual findings. First, the estimated $\text{NIE}(e)$, $\text{NDE}(e)$, and $\text{TE}(e)$ at exposure time $e = 6$ are negative. This phenomenon may be attributed to a potential treatment fatigue associated with the longest exposure duration, which could result in a reversal or attenuation of the treatment effect. Second, an inconsistent mediation result \citep{mackinnon2012introduction} was observed for $e=3$, leading to $\text{MP}(3)$ exceeding 1. This could be due to an increasingly smaller effective sample size for assessing mediation with a larger exposure time, under a stepped wedge design, but might also be attributed to potentially unmeasured mediator-outcome confounding. Third, the confidence intervals for the total effect and natural direct effect are the widest at $e=4$. Our further investigation shows that, during the jackknife procedure when we remove data in health facility 3, the estimate of $\theta_4$ in model (\ref{bina_Y_etm}) inflates dramatically, leading to inflated $\widehat{\text{NDE}}(4)$ and $\widehat{\text{TE}}(4)$ in $\hat{\bxi}_{-3}$, defined in (\ref{jackknife}). This could be because health facility 3 carries larger information content for estimating effects at $e=4$, a concept that was previously studied in the context of a constant treatment effect structure \citep{Kasza2019a,Mildenberger2023,Li2023}. For these reasons, it may be more informative (and the results are likely more numerically stable) to interpret summary mediation measures across exposure times, even when exposure time-dependent models are considered. Overall, the summary MP is estimated to be about 78.3\% (with the GHQ approach) and is comparable to the MP estimate under the constant treatment effect models.}

We further explore the potential for exposure time effect heterogeneity by testing equality of total effects over $e$, that is, 
$$H_0: \text{TE}(1|\bx^*)=\text{TE}(2|\bx^*)=\text{TE}(3|\bx^*)=\text{TE}(4|\bx^*)=\text{TE}(5|\bx^*)=\text{TE}(6|\bx^*),$$ 
where $\text{TE}(e|\bx^*)=\sum_{j=e+2}^J \text{TE}(j,e|\bx^*)/(J-e)$, and $\bx^*$ is the median level of all confounders.  
Under the null hypothesis, we consider a quadratic test statistic $\widehat{S}^T\widehat{\Sigma}^{-1}\widehat{S}$, where $\widehat{S}=(\widehat{\text{TE}}(1|\bx^*)- \widehat{\text{TE}}(2|\bx^*), \widehat{\text{TE}}(1|\bx^*)- \widehat{\text{TE}}(3|\bx^*), \widehat{\text{TE}}(1|\bx^*)- \widehat{\text{TE}}(4), 
\widehat{\text{TE}}(1|\bx^*)- \widehat{\text{TE}}(5|\bx^*), 
\widehat{\text{TE}}(1|\bx^*)- \widehat{\text{TE}}(6|\bx^*))^T$, and $\widehat{\Sigma}$ is estimated by jackknife under the exposure time-dependent models. We compare the test statistic against the asymptotic null distribution $\chi^2(5)$, and get a p-value of $0.44$ (with the GHQ approach). This may suggest that there is insufficient evidence from the observed data to support exposure time heterogeneity in the total effect, and hence the fluctuating patterns of the mediation effects under the exposure time-dependent analysis could have been due to random variation. 

\textcolor{black}{Finally, the above mediation analysis is also repeated using generalized linear mixed models with a nested exchangeable random-effects structure, by including an additional random effect at the cluster-period level for both the outcome model and the mediator model. The results are presented in Table S9 of Web Appendix F, and are generally similar to those in Table \ref{tab:res_data}. This may suggest that the random intercept models may have already been sufficient for analyzing this current data example.}

\section{Discussion}\label{sec:discussion}
In this article, we have contributed regression-based methods to conduct mediation analysis in a cross-sectional SW-CRT. We have operated under the generalized linear mixed modeling framework, which is currently the most widely accessible approach for analyzing SW-CRTs \citep{Li2021,wang2024achieve}. We have derived the mediation expressions under four different data types, allowing the outcome and mediator to be either continuous or binary.  
In particular, when the mediator is binary, we provide two approximate methods to estimate the mediation measures. When outcome is binary and mediator is continuous, we further propose the double STA method to estimate mediation measures, which can avoid computationally intensive double integration. Extensive simulations are conducted to evaluate the performances of $\widehat{\rm NIE}$, $\widehat{\rm NDE}$, $\widehat{\rm TE}$ and $\widehat{\rm MP}$ under various scenarios, and confirm that our proposed methods have small bias and nominal coverage, even when the number of clusters is $15$. Additional work is needed to further examine the performance of our methods with a smaller number of clusters, under different data types, and to develop suitable finite-sample corrections \citep{ouyang2024maintaining}. 
Finally, to facilitate the implementation of our methods in practice, we have provided an R package \texttt{mediateSWCRT}, available at \url{https://github.com/Zhiqiangcao/mediateSWCRT}. \textcolor{black}{We have also provided a short tutorial with simulated data sets in the Web Appendix E for demonstrating the key syntax. As we demonstrated in Web Appendix E and the data analysis in Section \ref{sec:application}, our approach can also accommodate incomplete designs with an implementation period, as the core functions depend on existing routines for fitting generalized linear mixed models. Nevertheless, this R package is tailored for mediation analysis for cross-sectional SW-CRTs, and hence does not address simpler parallel-arm design scenarios; the exact mediation expressions for other cluster randomized designs may require a separate development.}

As mediation analysis has received increasing attention in recent publications of either completed SW-CRTs  \citep{stevens2019mechanisms,gosselin2023immigrants} and protocols \citep{suresh2022pathweigh,shelley2015testing}, our goal is to contribute valid and accessible methods to address the unique complexity in such designs to improve practice. Although to the best of our knowledge this is the first article that addresses mediation analysis of cross-sectional SW-CRTs, our development is not exhaustive and there are several important future directions. First, 
while our primary focus is on cross-sectional SW-CRTs, closed-cohort SW-CRTs necessitate additional considerations for mediation analysis and need a separate development. \textcolor{black}{This is because, when the same individuals are repeated assessed over multiple periods, the mediator and outcome measured in previous periods can have a direct causal pathway to those measured in subsequent periods, and additional assumptions and estimands are required to tackle the clustered time-varying mediation data structure \citep{VanderWeele2015}. Therefore, simply including an additional individual-level random effects (as done in the conventional primary analysis of closed-cohort SW-CRTs) is not sufficient, and formal causal mediation methods with appropriate estimands and the requisite identification assumptions in closed-cohort SW-CRTs were introduced by \citet{yang2024sensitivity}}. Second, for simplicity, we have assumed that there are no interactions between exposure and mediator in the outcome model. Investigating the expressions for mediation effect measures and estimation methods with an exposure-mediator interaction in SW-CRTs is an important direction for future research \citep{vanderweele2016mediation}. 
\textcolor{black}{Third, while our development is based on the generalized linear mixed modeling framework, an alternative and equally valuable approach for analyzing cross-sectional SW-CRTs is the marginal modeling framework fitted using generalized estimating equations \citep{LiTurnerPreisser2018,li2022marginal,tian2022impact,li2023generalizing,zhang2023geemaee,zhang2023general,liu2024optimal}. Unlike mixed models, the marginal modeling framework separates the modeling of correlation structures from mean structures, potentially simplifying the mediation expressions by eliminating the need to integrate over random effects. The comparative strengths and limitations of generalized linear mixed modeling and marginal modeling frameworks for analyzing SW-CRTs were discussed in Section 5.2 of \citet{ouyang2024maintaining}. We plan to extend our software to support mediation analysis under marginal models in future work.}

\textcolor{black}{Finally, in the main manuscript, we have focused our development based on models with a random intercept in order to simplify the presentation. To provide some protection against potential misspecification of random-effects structure, we have adopted the cluster jackknife variance estimator rather than the model-based variance estimator. While the cluster jackknife variance estimator has been shown to maintain valid inference under misspecified linear mixed models \citep{ouyang2024maintaining}, its performance under misspecified generalized linear mixed models with binary outcomes warrants further simulation investigation. To advance our methods toward more complex random-effects structures, we present an extension of the mediation effect measures under a working nested exchangeable random-effects structure in Web Appendix C. This extension aligns with the recommendation to distinguish between within-period and between-period correlations in SW-CRTs \citep{taljaard2016substantial}. The corresponding mediation measure expressions are summarized in Table S1 of Web Appendix F. In comparison to the simpler models discussed in Sections \ref{sec:const_te} and \ref{sec:td_te}, estimating mediation effects for binary data under the nested exchangeable random-effects structure is computationally more demanding due to the necessity of integrating over additional random effects. Detailed approximation techniques are provided in Web Appendix C. In Section \ref{sec:application}, our additional sensitivity analyses under the nested exchangeable random-effects models provide similar results on point and standard error estimates to those in Table \ref{tab:res_data}, which may imply that the simpler random-effects structure is adequate for the specific SW-CRT. To increase model flexibility in practical applications, our \texttt{mediateSWCRT} R package supports the specification of both the simple exchangeable and the nested exchangeable random-effects structures. Extending our methodology to incorporate other random-effects structures, such as the exponential decay structure, remains an important area for future research \citep{Li2021,Kasza2019}}.

\section*{Acknowledgement}
Research in this article was supported by the Patient-Centered Outcomes Research Institute\textsuperscript{\textregistered} (PCORI\textsuperscript{\textregistered} Award ME-2023C1-31350). The statements presented in this article are solely the responsibility of the authors and do not necessarily represent the views of PCORI\textsuperscript{\textregistered}, its Board of Governors or Methodology Committee. The authors thank the participants and study team of the 99DOTS stepped wedge trial and Professor Adithya Cattamanchi for sharing the dataset. We also thank Hao Wang for helpful discussions with the data analysis.

\section*{Data Availability Statement}
Access policies for the original data set of the SW-CRT
in Uganda can be found in \citet{Cattamanchi2021} at \url{https://journals.plos.org/plosmedicine/article?id=10.1371/journal.pmed.1003628}; details
on the collection of that data can be found in \citet{Crowder2020} at \url{https://bmjopen.bmj.com/content/10/11/e039895}.

\section*{Supporting information}
Additional supporting information including Web Appendices A--E can be found at Github platform \url{https://github.com/Zhiqiangcao/mediateSWCRT}.

\bibliographystyle{jasa3}
\bibliography{DRGEN}

\clearpage

\clearpage
\begin{center}
\footnotesize
	\begin{tikzcd}[cells={nodes={draw}}, row sep=1cm, column sep=0.5cm]
		& & \textup{\shortstack{Clarify scientific question, and\\
				prepare data for mediation analysis}}\arrow[d] & \\
		& & \textup{\shortstack{Specify outcome, mediator, \\ treatment, cluster and period \\ variables in the \texttt{mediate\_swcrt}()\\ function}}\arrow[d] & \\
		& \textup{\shortstack{Is the mediator\\ binary?}}\arrow[dl, bend right=40, "Yes"']\arrow[d, "No"] & 
		\textup{\shortstack{Is the outcome\\ binary?}}\arrow[l, "Yes" ']\arrow[r, "No"] &
		\textup{\shortstack{Is the mediator\\ binary?}}\arrow[d, "Yes"']\arrow[dr, bend left=40, "No"] & \\
		\textup{\shortstack{$Y_bM_b$ type, set\\ \texttt{binary.o=1},\\ \texttt{binary.m=1}}}\arrow[drr] & 
		\textup{\shortstack{$Y_bM_c$ type, set\\ \texttt{binary.o=1},\\ \texttt{binary.m=0}}} \arrow[dr] & &  
		\textup{\shortstack{$Y_cM_b$ type, set\\ \texttt{binary.o=0},\\ \texttt{binary.m=1}}}\arrow[dl] & 
		\textup{\shortstack{$Y_cM_c$ type, set\\ \texttt{binary.o=0},\\ \texttt{binary.m=0}}}\arrow[dll]\\
		& & \textup{\shortstack{Specify \texttt{covariate.outcome}, \\
				and \texttt{covariate.mediator} in the\\
				\texttt{mediate\_swcrt}() function\\
				(default is \texttt{NULL})}}\arrow[d]\\
		& \textup{\shortstack{Set\\ \texttt{time.dependent}\\ \texttt{=TRUE}}} \arrow[dr, bend right = 10] & 
		\textup{\shortstack{Is the treatment structure\\ expected to be exposure-time\\ dependent?}}\arrow[l, "Yes"']\arrow[r, "No"] &
		\textup{\shortstack{Set\\ \texttt{time.dependent}\\ \texttt{=FALSE}}}\arrow[dl, bend left = 10]  & \\
		& & \textup{\shortstack{Output estimated mediation\\ effect measures and the \\95\% confidence intervals}}
	\end{tikzcd}
	\captionof{figure}{\textcolor{black}{A flow diagram with key decision points for implementing mediation analysis in cross-sectional SW-CRTs using the developed \texttt{mediateSWCRT} R package. Of note, the package also allows for the choice of a simple exchangeable working random-effects structure (developed in the main paper) or a nested exchangeable working random-effects structure (developed in Web Appendix C), which is not visualized in the diagram for brevity.}}
	\label{fig:workflow}
\end{center}

\clearpage
\begin{landscape}
\begin{table}
\begin{center}
\caption{Summary of expressions of mediation measures, period-specific $\text{NIE}(j|\bx)$ and $\text{NDE}(j|\bx)$ as well as calendar-exposure-time specific $\text{NIE}(j,e|\bx)$ and $\text{NDE}(j,e|\bx)$ under four data types in SW-CRTs. Under an instantaneous and constant treatment effect structure, the overall summary NIE and NDE measures can be obtained based on $\text{NIE}(j|\bx)$ and $\text{NDE}(j|\bx)$ by applying (\ref{ave_def_total}). The definitions for $\kappa(a,j|\bx)$, $\mu(a,a^*,j|\bx)$ and $\lambda(a,a^*,j|\bx)$ for $a,a^*\in \{0,1\}$ are given in \eqref{eq:kappa}, \eqref{eq:d-int} and \eqref{eq:lambda}, respectively. Under an exposure-time dependent treatment effect structure, the exposure-time specific $\text{NIE}(e|\bx)$ and $\text{NDE}(e|\bx)$ measures can be obtained based on $\text{NIE}(j,e|\bx)$ and $\text{NDE}(j,e|\bx)$ by applying (\ref{def_td}), and the overall summary NIE and NDE measures can be obtained by (\ref{def_td_total}) directly. The definitions for $\kappa_e(a,j|\bx)$, $\mu_e(a,a^*,j|\bx)$ and $\lambda_e(a,a^*,j|\bx)$ for $a,a^*\in \{0,1\}$ are given in \eqref{eq:kappa_etm}, \eqref{eq:mu_etm} and \eqref{eq:lambda_etm}, respectively.  
}
\begin{adjustbox}{width=1.3\textwidth}
\begin{tabular}{lllllcc}\label{tb:summary1}\\
	\toprule
	& & & \multicolumn{4}{c}{Instantaneous and constant treatment effect structure}\\
	\cmidrule{4-7}
	Data type & $Y$ & $M$ &  $\text{NIE}(j|\bx)$ & $\text{NDE}(j|\bx)$   & Depends on $\bx$? & Depends on $j$?\\	
	\midrule
	$Y_cM_c$ & Continuous & Continuous  & $\beta_{M}\eta $ & $\theta$ & $(\times,~\times)$ & $(\times,~\times)$\\
	$Y_cM_b$ & Continuous & Binary  & $\beta_{M}\left[\kappa(1,j|\bx)-\kappa(0,j|\bx)\right]$ & $\theta$ & $(\checkmark,~\times)$ & $(\checkmark,~\times)$\\
	$Y_bM_c$ & Binary & Continuous  & $\text{logit}\left[{\mu(1,1,j|\bx)}\right]-\text{logit}\left[{\mu(1,0,j|\bx)}\right]$ 
	&$\text{logit}\left[{\mu(1,0,j|\bx)}\right]-\text{logit}\left[{\mu(0,0,j|\bx)}\right]$ & $(\checkmark,~\checkmark)$ & $(\checkmark,~\checkmark)$\\
	$Y_bM_b$ & Binary & Binary  &  $\text{logit}\left[\lambda(1,0,j|\bx)[1-\kappa(1,j|\bx)]+\lambda(1,1,j|\bx)\kappa(1,j|\bx)\right]$ & $\text{logit}\left[\lambda(1,0,j|\bx)[1-\kappa(0,j|\bx)]+\lambda(1,1,j|\bx)\kappa(0,j|\bx)\right]$ & $(\checkmark,~\checkmark)$ & $(\checkmark,~\checkmark)$\\
	& & & $-\text{logit}\left[\lambda(1,0,j|\bx)[1-\kappa(0,j|\bx)]+\lambda(1,1,j|\bx)\kappa(0,j|\bx)\right]$ & $-\text{logit}\left[\lambda(0,0,j|\bx)[1-\kappa(0,j|\bx)]+\lambda(0,1,j|\bx)\kappa(0,j|\bx)\right]$ &\\
	\midrule
	& & & \multicolumn{4}{c}{Exposure-time dependent treatment effect structure (for a given exposure time level $e$)}\\
	\cmidrule{4-7}
	Data type & $Y$ & $M$ &  $\text{NIE}(j,e|\bx)$ & $\text{NDE}(j,e|\bx)$   & Depends on $\bx$? & Depends on $j$?\\	
	\midrule
	$Y_cM_c$ & Continuous & Continuous  & $\beta_{M}\eta_e $ & $\theta_e$ & $(\times,~\times)$ & $(\times,~\times)$\\
	$Y_cM_b$ & Continuous & Binary  & $\beta_{M}\left[\kappa_e(1,j|\bx)-\kappa_e(0,j|\bx)\right]$ & $\theta_e$ & $(\checkmark,~\times)$ & $(\checkmark,~\times)$\\
	$Y_bM_c$ & Binary & Continuous  & $\text{logit}\left[{\mu_e(1,1,j|\bx)}\right]-\text{logit}\left[{\mu_e(1,0,j|\bx)}\right]$ 
	&$\text{logit}\left[{\mu_e(1,0,j|\bx)}\right]-\text{logit}\left[{\mu_e(0,0,j|\bx)}\right]$ & $(\checkmark,~\checkmark)$ & $(\checkmark,~\checkmark)$\\
	$Y_bM_b$ & Binary & Binary  &  $\text{logit}\left[\lambda_e(1,0,j|\bx)[1-\kappa_e(1,j|\bx)]+\lambda_e(1,1,j|\bx)\kappa_e(1,j|\bx)\right]$ & $\text{logit}\left[\lambda_e(1,0,j|\bx)[1-\kappa_e(0,j|\bx)]+\lambda_e(1,1,j|\bx)\kappa_e(0,j|\bx)\right]$ & $(\checkmark,~\checkmark)$ & $(\checkmark,~\checkmark)$\\
	& & & $-\text{logit}\left[\lambda_e(1,0,j|\bx)[1-\kappa_e(0,j|\bx)]+\lambda_e(1,1,j|\bx)\kappa_e(0,j|\bx)\right]$ & $-\text{logit}\left[\lambda_e(0,0,j|\bx)[1-\kappa_e(0,j|\bx)]+\lambda_e(0,1,j|\bx)\kappa_e(0,j|\bx)\right]$ &\\
	\bottomrule			
\end{tabular}
\end{adjustbox}
\end{center}
\end{table} 
\end{landscape}
\clearpage

\begin{table}
\begin{center}
\caption{Simulation results for mediation measures under four data types in SW-CRTs with an instantaneous and constant treatment effect structure under different number of clusters. GHQ: Gauss-Hermite Quadrature approach for single integral calculation; STA: second-order Taylor approximation for single and double integral calculation.}\label{tab:sim1}
\begin{adjustbox}{width=1\textwidth}
			\begin{tabular}{clcrrrrrrrrrrrrrr}
				\hline
				& & &\multicolumn{4}{c}{$I=15$} & &\multicolumn{4}{c}{$I=30$}& &\multicolumn{4}{c}{$I=60$} \\
				\cline{4-7}\cline{9-12}\cline{14-17}  Datatype &Parameter  & TRUE & Bias(\%) & MCSD & AESE & CP(\%) & & Bias(\%) & MCSD & AESE & CP(\%)  & & Bias(\%) & MCSD & AESE & CP(\%)  \\
				\hline
				$Y_cM_c$ &{\rm NIE} &0.25 & -0.28 & 0.07 & 0.07 & 95.4 && -0.06 & 0.05 & 0.05 & 95.4 && -0.13 & 0.03 & 0.03 & 95.6\\
				&{\rm NDE} &0.75 & -0.07 & 0.11 & 0.11 & 96.1 &&  0.01 & 0.07 & 0.08 & 95.3 && -0.03 & 0.05 & 0.05 & 95.4 \\ 
				&{\rm TE} & 1.00 & -0.34 & 0.13 & 0.13 & 95.4 && -0.05 & 0.09 & 0.09 & 95.7 && -0.16 & 0.06 & 0.06 & 96.6 \\ 
				&{\rm MP} & 0.25 & -0.28 & 0.06 & 0.06 & 96.5 && -0.08 & 0.04 & 0.04 & 95.8 && -0.10 & 0.03 & 0.03 & 95.0 \\
				$Y_cM_b$ &\underline{GHQ} && && && && && && && &\\
				&{\rm NIE} &0.25 & 0.68 & 0.14 & 0.15 & 95.7 && -0.03 & 0.10 & 0.10 & 95.6 && -0.27 & 0.07 & 0.07 & 95.5 \\ 
				&{\rm NDE} & 0.75 & 0.01 & 0.10 & 0.11 & 95.4 && 0.05 & 0.08 & 0.08 & 94.2 && 0.07 & 0.05 & 0.05 & 96.0 \\ 
				&{\rm TE} & 1.00 & 0.69 & 0.18 & 0.19 & 96.3 && 0.02 & 0.13 & 0.13 & 95.1 && -0.20 & 0.09 & 0.09 & 94.7 \\ 
				&{\rm MP} & 0.25 & -0.82 & 0.12 & 0.12 & 95.7 && -0.74 & 0.08 & 0.08 & 95.6 && -0.56 & 0.05 & 0.06 & 95.8 \\
				&\underline{STA} && && && && && && && &\\
				&{\rm NIE} &0.25 & 0.48 & 0.14 & 0.15 & 95.7 && -0.23 & 0.10 & 0.10 & 95.6 && -0.47 & 0.07 & 0.07 & 95.7 \\ 
				&{\rm NDE} & 0.75 & 0.01 & 0.10 & 0.11 & 95.4 && 0.05 & 0.08 & 0.08 & 94.2 && 0.07 & 0.05 & 0.05 & 96.0 \\ 
				&{\rm TE} & 1.00 & 0.50 & 0.17 & 0.19 & 96.5 && -0.17 & 0.13 & 0.13 & 95.0 && -0.40 & 0.09 & 0.09 & 94.9 \\ 
				&{\rm MP} & 0.25 & -0.94 & 0.12 & 0.12 & 95.8 && -0.88 & 0.08 & 0.08 & 95.9 && -0.71 & 0.05 & 0.06 & 95.7 \\
				$Y_bM_c$ &\underline{STA} && && && && && && && &\\
				&{\rm NIE} &0.25 & 0.37 & 0.07 & 0.07 & 95.8 && 0.22 & 0.05 & 0.05 & 95.3 && 0.06 & 0.03 & 0.04 & 96.5\\
				&{\rm NDE} &0.75 & 1.47 & 0.22 & 0.23 & 96.4 && -0.41 & 0.15 & 0.15 & 95.1 && -0.11 & 0.11 & 0.11 & 94.9\\ 
				&{\rm TE} & 1.00 & 1.84 & 0.23 & 0.24 & 96.1 && -0.19 & 0.16 & 0.16 & 95.0 && -0.05 & 0.11 & 0.11 & 95.3\\ 
				&{\rm MP} & 0.25 & 0.85 & 0.09 & 0.09 & 95.3 && 0.69 & 0.06 & 0.06 & 96.1 && 0.28 & 0.04 & 0.04 & 95.7 \\
				$Y_bM_b$ &\underline{GHQ} && && && && && && && &\\
				&{\rm NIE} &0.25 & -0.01 & 0.14 & 0.15 & 96.8 && 0.02 & 0.10 & 0.10 & 95.3 && -0.21 & 0.07 & 0.07 & 95.8\\ 
				&{\rm NDE} & 0.75 & 2.42 & 0.29 & 0.29 & 95.3 && 1.08 & 0.20 & 0.20 & 95.7 && 0.25 & 0.13 & 0.13 & 95.8 \\ 
				&{\rm TE} & 1.00 & 2.41 & 0.32 & 0.33 & 96.2 && 1.11 & 0.22 & 0.22 & 95.6 && 0.03 & 0.15 & 0.15 & 95.7\\ 
				&{\rm MP} & 0.25 & 0.15 & 0.16 & 0.21 & 98.1 && -0.03 & 0.10 & 0.10 & 97.2 && -0.16 & 0.07 & 0.07 & 95.9 \\
				&\underline{STA} && && && && && && && &\\
				&{\rm NIE} &0.25 & -0.24 & 0.14 & 0.15 & 96.9 && -0.22 & 0.10 & 0.10 & 95.2 && -0.46 & 0.07 & 0.07 &95.6 \\ 
				&{\rm NDE} &0.75 & 1.78 & 0.29 & 0.29 & 95.3 && 0.45 & 0.20 & 0.19 & 95.5 && -0.39 & 0.13 & 0.13 & 96.0 \\ 
				&{\rm TE}  &1.00 & 1.54 & 0.32 & 0.33 & 96.4 && 0.23 & 0.22 & 0.22 & 95.4 && -0.85 & 0.15 & 0.15 & 95.6 \\ 
				&{\rm MP}  &0.25 & 0.15 & 0.16 & 0.31 & 98.1 && -0.05 & 0.10 & 0.10 & 96.8 && -0.18 & 0.07 & 0.07 &96.0 \\  
				\hline
			\end{tabular}
		\end{adjustbox}
	\end{center}
\end{table}

\clearpage

\begin{table}
\tiny
	\begin{center}
		\caption{Mediation analysis results in the 99DOTS-based SW-CRT. The NIE, NDE, TE and MP are all defined on a log odds ratio scale, conditional on the median level of the covariates. S.E. denotes the standard error estimate calculated by jackknife. The notation $^*$ indicates that the lower boundary or upper boundary of CI for MP is non-informative since MP should be strictly bounded between 0 and 1. In practice, one can winzorize the interval to be within $[0,1]$ for ease of interpretation.}\label{tab:res_data}
		\begin{tabular}{llrrcrrrc}
			\hline
			& & \multicolumn{3}{c}{GHQ} & &\multicolumn{3}{c}{STA}  \\
			\cline{3-5}\cline{7-9}  Treatment effect &Parameter  & Estimate & S.E. & 95\% CI &  & Estimate & S.E. & 95\% CI  \\
			\hline
			Constant& {\rm NIE} &1.059&0.340&(0.341,1.776)&&1.059& 0.341 &(0.340,1.779)\\ 
			& {\rm NDE} &0.255 & 0.459 &(-0.713,1.224)&&0.256 & 0.459 &(-0.714, 1.225)\\   
			& {\rm TE}  &1.314 & 0.543 &(0.168,2.460)&&1.315 & 0.544 &(0.167,2.463) \\  
			& {\rm MP}  &0.806 & 0.291 &(0.192,1.419)$^*$&&0.806 & 0.291 &(0.192,1.419)$^*$ \\  
			Time-dependent  &\underline{$e=1$}& &&&&&&\\
			&{\rm NIE}($e$) &0.991 & 0.410 &(0.126,1.856)&&0.992 & 0.411 &(0.125,1.859) \\ 
			&{\rm NDE}($e$) &0.151 & 0.582 &(-1.077,1.379)&&0.151&0.583&(-1.079,1.381) \\ 
			&{\rm TE}($e$)  &1.142 & 0.667 &(-0.266,2.550)&&1.143&0.668&(-0.267,2.553) \\ 
			&{\rm MP}($e$)  &0.868 & 0.446 &(-0.074,1.810)$^*$&&0.868&0.447&(-0.075,1.811)$^*$ \\ 
			&\underline{$e=2$}& &&&&&&\\
			&{\rm NIE}($e$) &1.226 & 0.420 &(0.339,2.113)&&1.227&0.422 &(0.336,2.118) \\ 
			&{\rm NDE}($e$) &0.576 & 0.758 &(-1.022,2.175)&&0.579&0.761&(-1.027,2.185) \\ 
			&{\rm TE}($e$)  &1.802 & 0.939 &(-0.180,3.784)&&1.806&0.944&(-0.187,3.799) \\
			&{\rm MP}($e$)  &0.680 & 0.276 &(0.098,1.263)$^*$&&0.679&0.276 &(0.097,1.262)$^*$ \\  
			&\underline{$e=3$}& &&&&&&\\
			&{\rm NIE}($e$) &1.412 & 0.669 &(0.001,2.824)&&1.414& 0.671 &(-0.002,2.831) \\ 
			&{\rm NDE}($e$) &-0.187& 0.750 &(-1.769,1.396)&&-0.187&0.752&(-1.774,1.400) \\ 
			&{\rm TE}($e$)  &1.226 & 0.819 &(-0.502,2.954)&&1.227&0.823&(-0.509,2.964) \\
			&{\rm MP}($e$)  &1.152 & 0.731 &(-0.390,2.694)$^*$&&1.152&0.733&(-0.395,2.699)$^*$ \\ 
			&\underline{$e=4$}& &&&&&&\\
			&{\rm NIE}($e$) &0.572 & 0.717&(-0.940,2.084)&&0.572 &0.718&(-0.942,2.086) \\ 
			&{\rm NDE}($e$) &1.273 &14.736&(-29.818,32.363)&&1.281&14.753&(-29.845,32.407) \\
			&{\rm TE}($e$)  &1.844 &14.589&(-28.937,32.625)&&1.853&14.606&(-28.964,32.669) \\
			&{\rm MP}($e$)  &0.310 &0.388 &(-0.509,1.129)$^*$&& 0.309&0.387&(-0.507,1.125)$^*$ \\  
			&\underline{$e=5$}& &&&&&&\\
			&{\rm NIE}($e$) &1.616 & 0.815 &(-0.104,3.336)&&1.618&0.817&(-0.106,3.342) \\  
			&{\rm NDE}($e$) &0.247 & 1.080 &(-2.032,2.526)&&0.248&1.087&(-2.045,2.542) \\ 
			&{\rm TE}($e$)  &1.863 & 1.394 &(-1.077,4.803)&&1.867&1.401&(-1.089,4.822) \\
			&{\rm MP}($e$)  &0.867 & 0.612 &(-0.424,2.159)$^*$&&0.867&0.616&(-0.432,2.166)$^*$ \\ 
			&\underline{$e=6$}& &&&&&&\\
			&{\rm NIE}($e$) &-0.198&1.061&(-2.436,2.040)&&-0.198 & 1.062 &(-2.439,2.042) \\
			&{\rm NDE}($e$) &-0.501&0.932&(-2.467,1.465)&&-0.500 & 0.935 &(-2.472,1.471) \\ 
			&{\rm TE}($e$)  &-0.700&1.581&(-4.035,2.636)&&-0.699 & 1.582 &(-4.037,2.639) \\
			&{\rm MP}($e$)  &0.283&1.172 &(-2.189,2.756)$^*$&& 0.284 & 1.173 &(-2.190,2.758)$^*$ \\ 
			&\underline{overall summary}& &&&&&&\\
			&{\rm NIE} &0.936 & 0.397 &(0.100,1.773) & & 0.937 & 0.398 &(0.098,1.777) \\  
			&{\rm NDE} &0.260 & 2.673 &(-5.379,5.899)& & 0.262 & 2.676 &(-5.385,5.909) \\  
			&{\rm TE}  &1.196 & 2.633 &(-4.358,6.750)& & 1.199 & 2.636 &(-4.363,6.762) \\   
			&{\rm MP}  &0.783 & 0.720 &(-0.736,2.302)$^*$& & 0.782 & 0.719 &(-0.736,2.299)$^*$ \\
			\hline			
		\end{tabular}
	\end{center}
\end{table}

\end{document}